\documentclass[twocolumn,times,tighten]{aastex631_mod}

\usepackage{newtxtext,newtxmath}
\usepackage[T1]{fontenc}
\usepackage{ae,aecompl}
\usepackage{braket}
\usepackage{etoolbox}
\makeatletter
\patchcmd\@combinedblfloats{\box\@outputbox}{\unvbox\@outputbox}{}{
}
\makeatother

\usepackage{graphicx}	% Including figure files
\usepackage{amsmath}	% Advanced maths commands
\usepackage{amssymb}	% Extra maths symbols
\usepackage{float}
\usepackage{verbatim}

\newcommand{\alpro}{\textsc{ALPro}}
\newcommand{\invgev}{$\,{\rm GeV}^{-1}$}
\newcommand{\g}{g_{a\gamma}}
\newcommand{\gmath}{$g_{a\gamma}$}
\newcommand{\lmin}{\Lambda_{\rm min}}
\newcommand{\lmax}{\Lambda_{\rm max}}
\newcommand{\lc}{\Lambda_{\rm c}}
\newcommand{\dz}{\delta z}
\newcommand{\pgg}{P_{\gamma \gamma}}

\newcommand{\pga}{P_{\gamma a}}
\newcommand{\pgax}{P_{\gamma_x \to a}}
\newcommand{\pgay}{P_{\gamma_y \to a}}
\newcommand{\omegap}{\omega_{\rm pl}}
\newcommand{\betapl}{\beta_{\rm pl}}
\newcommand{\Deltap}{\Delta_{\rm pl}}
\newcommand{\myvector}{\boldsymbol}
\newcommand{\Acal}{{\cal A}}
\newcommand{\xspec}{\textsc{xspec}}
\definecolor{C0}{HTML}{1f77b4}
\definecolor{C1}{HTML}{ff7f0e}
\definecolor{C2}{HTML}{2ca02c}
\definecolor{C3}{HTML}{d62728}
\definecolor{C6}{HTML}{e377c2}
\shorttitle{How do Magnetic Field Models Affect Astrophysical Limits on Light ALPs?}
\shortauthors{J. H. Matthews et al.}

\begin{document}
\title
{
How do Magnetic Field Models Affect Astrophysical Limits on Light Axion-like Particles? An X-ray Case Study with NGC 1275
}

% \correspondingauthor{James Matthews}
% \email{matthews@ast.cam.ac.uk}

% The list of authors, and the short list which is used in the headers.
% If you need two or more lines of authors, add an extra line using \newauthor
\newcommand{\ioa}{Institute of Astronomy, University of Cambridge, Madingley Road, Cambridge, CB3 0HA; \href{mailto:matthews@ast.cam.ac.uk}{matthews@ast.cam.ac.uk}}

\author[0000-0002-3493-7737]{James~H.~Matthews}
\affiliation{\ioa}

\author[0000-0002-1510-4860]{Christopher S. Reynolds}
\affiliation{\ioa}

\author[0000-0001-7271-4115]{M. C. David Marsh}
\affiliation{The Oskar Klein Centre, Department of Physics, Stockholm University, Stockholm 106 91, Sweden}

\author[0000-0003-3814-6796]{Júlia Sisk-Reynés}
\affiliation{\ioa}

\author[0000-0002-1624-9359]{Payton E. Rodman}
\affiliation{\ioa}

\received{2022 Jan 19}
\revised{2022 Feb 15}
\accepted{2022 Feb 17}

% Abstract of the paper
\begin{abstract}
\noindent
Axion-like particles (ALPs) are a well-motivated extension to the standard model of particle physics, and X-ray observations of cluster-hosted AGN currently place the most stringent constraints on the ALP coupling to electromagnetism, $\g$, for very light ALPs ($m_a\lesssim10^{-11}$\,eV). We revisit limits obtained by \cite{reynolds_astrophysical_2020} using {\sl Chandra} X-ray grating spectroscopy of NGC 1275, the central AGN in the Perseus cluster, examining the impact of the X-ray spectral model and magnetic field model. We also present a new publicly available code, \textsc{ALPro}, which we use to solve the ALP propagation problem. We discuss evidence for turbulent magnetic fields in Perseus and show that it can be important to resolve the magnetic field structure on scales below the coherence length. We re-analyse the NGC 1275 X-ray spectra using an improved data reduction and baseline spectral model. We find the limits are insensitive to whether a partially covering absorber is used in the fits. At low $m_a$ ($m_a\lesssim10^{-13}$\,eV), we find marginally weaker limits on $\g$ (by $0.1-0.3$\,dex) with different magnetic field models, compared to Model B from \cite{reynolds_astrophysical_2020}. A Gaussian random field (GRF) model designed to mimic $\sim50$\,kpc scale coherent structures also results in only slightly weaker limits. We conclude that the existing Model B limits are robust assuming that $\betapl\approx100$, and are insensitive to whether cell-based or GRF methods are used. However, astrophysical uncertainties regarding the strength and structure of cluster magnetic fields persist, motivating high sensitivity RM observations and tighter constraints on the radial profile of $\betapl$.
\end{abstract}

%%%%%%%%%%%%%%%%% BODY OF PAPER %%%%%%%%%%%%%%%%%%

\section{Introduction}
Probing physics beyond the standard model (SM) of particle physics is a fundamental goal of modern particle and astroparticle physics. One particularly well-motivated SM extension involves axions. The axion is the particle associated with the Peccei-Quinn field \citep{peccei_cp_1977,weinberg_new_1978,wilczek_problem_1978}, and was posited to solve the strong CP problem of Quantum Chromodynamics \citep[QCD;][]{cheng_strong_1988,kim_axions_2010}. Including the QCD axion field naturally leads to CP conservation without a fine-tuning problem. The QCD axion has a one-to-one relationship between its mass, $m_a$, and coupling to electromagnetism, \gmath, but a more general class of particles known as axion-like particles \citep[ALPs; see reviews by][]{graham_experimental_2015,irastorza_new_2018} is predicted by effective theories derived from string theory \citep{svrcek_axions_2006}. ALPs are appealing dark matter candidates \citep{preskill_cosmology_1983,ringwald_exploring_2012,arias_wispy_2012,marsh_axion_2016,chadha-day_axion_2021} and can modify astrophysical processes \cite[e.g.][]{raffelt_stars_1996}; their significance for fundamental particle physics, astrophysics and cosmology is therefore clear.

An important characteristic of ALPs is their coupling to radiation with a mass-independent coupling constant, described by the Lagrangian term 
\begin{equation}
{\cal L}_{\gamma a} = \g a (\boldsymbol{E} \cdot \boldsymbol{B}),
\end{equation}
where $a$ is the ALP field, $\boldsymbol{E}$ is the electric field and $\boldsymbol{B}$ is the magnetic field. ALPs couple to two photons, meaning that in the presence of an external magnetic field, ALPs and photons can undergo quantum mechanical oscillations. The property of photon-ALP conversion or `mixing' in external magnetic fields can be used to search for ALPs experimentally, and place limits in $(m_a, \g)$ parameter space. Experimental ALP searches are reviewed by \cite{graham_experimental_2015}; approaches include so-called `light shining through wall' experiments \citep[e.g.][]{arias_optimizing_2010,ehret_resonant_2009,ballou_latest_2014} and axion helioscopes such as the CERN Axion Solar Telescope \citep[CAST;][]{cast_collaboration_improved_2007,arik_probing_2009} and the proposed International AXion Observatory \citep[IAXO;][]{irastorza_towards_2011,armengaud_conceptual_2014}. The presence of {\em astrophysical} magnetic fields can also be leveraged to search for ALPs. As an example, supernova 1987A provides limits on the photon-ALP coupling from the absence of an associated gamma-ray burst \citep[e.g.][]{brockway_sn_1996,raffelt_astrophysical_2008,payez_revisiting_2015}. Currently, one of the best ways of searching for and constraining light ALPs ($m_a \lesssim 10^{-9}$\,eV) involves searching for irregularities in the spectra of active galactic nuclei (AGN) embedded in magnetised clusters; this method, when applied to the gamma-ray and (in particular) X-ray frequency ranges, is our main focus here. 

\defcitealias{reynolds_astrophysical_2020}{R20}
\cite{wouters_constraints_2013} were the first to place constraints on ALPs from X-ray spectroscopy, using {\sl Chandra} data from Hydra A. Since then, ALP limits have been acquired from X-ray observations of M87 \citep{marsh_new_2017}, NGC 3862 \citep{conlon_constraints_2017}, and NGC 1275, the central AGN in the Perseus cluster \citep[][hereafter \citetalias{reynolds_astrophysical_2020}]{berg_constraints_2017,reynolds_astrophysical_2020}. \citetalias{reynolds_astrophysical_2020} found constraints on very light ALPs ($m_a \lesssim 10^{-11}$\,eV), ruling out $\g > 6-8 \times10^{-13}$\invgev\ at $99.7$ per cent confidence. Marginally tighter constraints still are obtained based on an analysis of the cluster-hosted quasar H1821$+$643 \citep{sisk_reynes_new_2021}.  \cite{schallmoser_updated_2021} find similar constraints from five AGN located either within or behind clusters such as Coma, using machine learning techniques as suggested by \cite{day_accelerating2020}. Gamma-ray studies offer slightly weaker, although complementary, constraints at higher masses, $5\times10^{-10} \lesssim (m_a/{\rm eV}) \lesssim 5\times10^{-9}$, with \cite{ajello_search_2016} excluding $\g\ > 5 \times10^{-12}$\invgev\ at $95$ per cent confidence. 

In the X-ray band, AGN are characterised by a power-law spectrum thought to be produced by inverse Compton scattering of accretion disc seed photons. This power-law spectrum is ubiquitous, but other spectral imprints such as a soft X-ray excess, relativistic reflection signatures and atomic features such as a $6.4$keV Iron line are also extremely common. These astrophysical signatures complicate searches for ALPs, since identifying any spectral irregularity relies on a well-characterised continuum source. While the underlying physics of accretion discs and their associated X-ray coronae, winds and absorbers is complex and poorly understood, from a phenomenological perspective AGN spectra can nevertheless be well modelled using a suite of detailed spectral models in tools such as \xspec\ \citep{arnaud_xspec_1996}. Furthermore, in some cases, the power-law is virtually featureless and only simple corrections for intervening soft X-ray absorption are needed, such as in the case of NGC 1275 in the Perseus cluster. However, additional complications stem from instrument calibration, photon pileup and the need to separate the cluster and AGN source \citepalias{reynolds_astrophysical_2020}.

A major systematic uncertainty in modelling ALP signatures from cluster AGN is the magnetic field structure and strength along our line of sight. This topic is the main focus of our work. The intracluster medium (ICM) is magnetised, and is likely to be turbulent \citep[e.g.][]{carilli_cluster_2002,govoni_magnetic_2004,schekochihin_turbulence_2006,vazza_resolved_2018,donnert_magnetic_2018}. Generally speaking, observations are consistent with a tangled magnetic field of $\sim 1-10 \mu {\rm G}$ strength, decreasing with radius, and a Kolmogorov power spectrum. The turbulence requires energy input, which could come from mergers and/or the central AGN. These processes may also create coherent large-scale fields, but evidence for such structures is relatively weak. Empirical constraints on the field strength can come from Faraday rotation, observed pressure profiles, synchrotron radio haloes and other model-dependent methods. Each of these has strengths and weaknesses. The quantity that matters for ALP conversion is the perpendicular magnetic field component ($B_\perp$) along the line of sight to the continuum source. The Faraday rotation measure (RM) gives us a line of sight measure, but it is an integrated quantity and only includes $B_\parallel$. The pressure profile does give an indication of the radial profile, but this is often measured within annuli around the source, so only corresponds to an accurate line of sight measure if a reasonable degree of spherical symmetry applies. 

This paper is structured as follows. First, in section~\ref{sec:alps}, we discuss the physics of ALP-photon conversion and introduce our new Python package, \alpro. We then review common models for modelling magnetic fields in clusters in ALP searches in section~\ref{sec:icm}, and explore the astrophysical evidence for magnetic fields in the Perseus cluster in particular. In section~\ref{sec:sensitivity}, we conduct a sensitivity study, comparing the results from Gaussian random fields and simpler cell-based approaches and assessing the impact of small-scale field structure. In section~\ref{sec:reanalysis}, we re-analyse the X-ray data from NGC 1275 in the Perseus cluster and present updated limits using various magnetic field models and an improved spectral model. In section~\ref{sec:discuss}, we discuss the application of the Fourier formalism to NGC 1275 as well as the implications for other clusters, before concluding in section~\ref{sec:conclusions}. Overall, we find that the astrophysical assumptions about the normalisation of the magnetic field strength are important and can modify the limits appreciably while still producing acceptable predicted Faraday rotation measures; however, other specific choices -- between a cell-based or Gaussian random field approach, or of how to set the coherence length of the magnetic field -- have a relatively small effect on the limits obtained, introducing a systematic uncertainty of $\sim 0.1$\,dex.

\section{ALP-Photon Interconversion}
\label{sec:alps}
The contribution to the Lagrangian from ALPs of mass $m_a$ can be written as
\begin{equation}
    {\cal L}_a = -\frac{1}{2} \partial_\mu a  \partial^\mu a -\frac{1}{2} m_a^2 a^2 + \g a (\boldsymbol{E} \cdot \boldsymbol{B})\, ,
\end{equation}
where the final term describes the ALP-photon mixing with coupling constant $\g$. We deal with relativistic ALPs, with $m_a \ll E$ such that the relevant equation of motion (for propagation in the $z$-direction) for a beam energy $E$ is a first-order Schr{\" o}dinger-like equation,  given by
\begin{equation}
	\left( i \frac{d}{d z} + E + {\cal M}(z) \right) \, 	
	\left( \begin{matrix}
	\Ket{\gamma_x} \\
	\Ket{\gamma_y} \\
	\Ket{a}
	\end{matrix} \right)
	= 0\, ,
\label{eq:schrodinger}
\end{equation}
where ${\cal M}$ is the mixing matrix of the form \citep{raffelt_mixing_1988}
\begin{equation}
	{\cal M}(z) = 
	\left( \begin{matrix}
	\Deltap(z) & 0 & \Delta_x(z) \\
	0 & \Deltap(z) & \Delta_y(z) \\
	\Delta_x(z) & \Delta_y(z)& \Delta_{a} \\
	\end{matrix} \right)\, .
\label{eq:mixing}
\end{equation}
Here the dispersive diagonal terms are $\Delta_{a} = -m_a^2/(2 E)$ and $\Deltap(z) = \omegap^2/(2E)$, where $\omegap$ is the usual plasma frequency. We have neglected the Faraday rotation terms, which are negligible at X-ray and gamma-ray energies. The off-diagonal terms are responsible for photon-ALP mixing and are given by 
\begin{align}
\Delta_x &= \g B_x(z)/2\,, \\
\Delta_y &= \g B_y(z)/2\, .
\label{eq:delta_y}
\end{align}
The Schr{\" o}dinger-like equation must in general be solved numerically, but analytical calculations are possible for certain configurations, and perturbative treatments are also useful (see, e.g, section~\ref{sec:fourier}). More mathematical details are given by other authors \citep[e.g.][]{raffelt_mixing_1988, de_angelis_relevance_2011,marsh_new_2017,davies_relevance_2020}; here we focus on discussing just a few key aspects of the ALP-photon conversion process. 

It is informative to consider an idealised case, where the beam is in a pure polarization state, $(\Ket{\gamma_x},\Ket{\gamma_y}, \Ket{a})=(1,0,0)$, and travels a distance $L$ through a uniform magnetic field of strength $B$ aligned with the $x$-axis.  In this case the off-diagonal $\Delta_y$ terms are zero and the conversion probability can be shown to be 
\begin{equation}
	P_{\gamma_{x}\rightarrow a} = \frac{\Theta^2}{1+\Theta^2} \sin^2 \left( \Delta_{\rm eff} \sqrt{1+\Theta^2} \right),
\end{equation}
where we have adopted the notation of \cite{marsh_new_2017} with $\Theta=2B_\perp E \g/m_{\rm eff}^2$ and $\Delta_{\rm eff}=m_{\rm eff}^2 L / (4E)$, where $m_{\rm eff}^2=m_a^2-\omegap^2$. Here we have given the conversion probability for a polarized beam, but in our work we assume that the X-ray emission is initially unpolarized. The degree of polarization of X-ray emission in AGN is not well constrained observationally -- although this may change with the launch of the {\sl Imaging X-ray Polarimetry Explorer [IXPE]}, as shown by, e.g., \cite{ursini2022} -- and the situation is complicated by the possibility of a composite X-ray source in NGC~1275 \citep[][see also section~\ref{sec:reanalysis}]{reynolds_probing_2021}. Models of X-ray polarization signatures from AGN predict an (inclination-dependent) degree of polarization of a few per cent from an accretion disc corona \citep{schnittman2010,beheshtipour2017} and $\sim10$ per cent from a relativistic jet \citep{mcnamara2009}. Given the absence of X-ray polarization data and the lack of knowledge about the detailed physics of the X-ray emitting region(s), we take a standard approach and do not assign any initial preferential polarization to the X-ray beam in our calculation. Additionally, we note that ALPs themselves can introduce polarization -- an effect that has been studied by \cite{day2018}, using NGC~1275 as a candidate source -- which has exciting prospects for future X-ray missions such as {\sl IXPE}. 

Under our assumptions, the actual multiplicative imprint on the spectrum is determined by the survival probability from an unpolarized beam, $\pgg$. This unpolarized survival probability can be calculated from the pure polarization case using 
\begin{equation}
	\pgg = (1 - \pga) = 1 - \frac{1}{2} \left( P_{\gamma_{x}\rightarrow a} + P_{\gamma_{y}\rightarrow a} \right),
\label{eq:unpol}
\end{equation}
where $\pga$ is the total, unpolarized conversion probability. In our work we will always consider the general case where both $\Delta_x$ and $\Delta_y$ are non-zero and $z$-dependent. In this case $\pgg$ must be calculated numerically by formulating a transfer matrix to solve equation~\ref{eq:schrodinger}. The line-of-sight is split into a series of cells and the calculation is carried out in a piecewise fashion in each cell $j$, with the output state $(\Ket{\gamma_x},\Ket{\gamma_y}, \Ket{a})$ used as input to the next cell $j+1$.  

\subsection{The Fourier formalism}
\label{sec:fourier}
Recently, \cite{marsh_fourier_2022} outlined a new formalism for treating relativistic ALP-photon conversion. \cite{marsh_fourier_2022} showed that, to leading order in $\g$, the conversion probability $P_{\gamma_i\to a}$ (with $i\in [x,y]$) can be related to $\Delta_i(z)$, or equivalently, the magnetic field profile along the line of sight $B_i(z)$, using Fourier-like transforms. Although this treatment breaks down when the conversion probabilities exceed $\sim 5$--$10$ per cent, it is an extremely useful framework when considering how different magnetic field treatments affect the conversion probability, so we briefly review the main results. In the massive ALP ($m_a \gg \omegap$) case and focusing on the $x$-component only, the conversion probability can be written as 
\begin{equation}
P_{\gamma_x \to a}(\eta) = {\cal F}_s( \Delta_{x})^2 + {\cal F}_c( \Delta_{x})^2 \\
\label{eq:prob_fourier1}
\end{equation}
where ${\cal F}_c$ and ${\cal F}_s$ denote the cosine and sine transforms using a conjugate variable $\eta=m_a^2/2E$, e.g. ${\cal F}_c(f) =\int^\infty_0 f(z) \cos(\eta z) dz$. By applying the Wiener-Khinchin theorem, the conversion probability can also be expressed in terms of a cosine transform of the autocorrelation function of the line-of-sight magnetic field, $c_{B_x}$, as 
\begin{equation}
P_{\gamma_x \to a}(\eta) = \frac{\g^2}{2} {\cal F}_c \Big( c_{B_x}(L)\Big).
\label{eq:prob_fourier2}
\end{equation}
In the massless case ($m_a \ll \omegap$), the same formalism applies if we transform to new variables.  Specifically, $\Delta_x$ is replaced by the function $G=2 \Delta_x/\omegap^2$, the line of sight distance coordinate is replaced by a phase factor proportional to the electron column density
\begin{equation}
    \varphi=\frac{1}{2}\int_0^z{\rm d}z^\prime\omegap^2(z^\prime),
\end{equation}
and the conjugate Fourier variable is $\lambda=1/E$. With these transformations,
%$\Delta_x$ is replaced by the function $G=2 \Delta_x/\omegap$ with a coordinate change from $z$ to a phase $\varphi$, and a conjugate variable  $\lambda=1/E$. Nevertheless, 
the basic principle is similar to the massive ALP case, as the conversion probability can still be expressed as a simple transform of a function of the line of sight perpendicular magnetic field. Although the forms above are for a polarized beam, the unpolarized survival probability can always be obtained directly from equation~\ref{eq:unpol}. We will discuss the applicability of this Fourier formalism to grid calculations for X-ray spectral fitting in section~\ref{sec:performance}.

\subsection{The \alpro\ Python Package}
\label{sec:alpro}
We use our new Python package \alpro\  (Axion-Like PROpagation; \citealt{alpro}), v1.0, to solve the Schr{\" o}dinger-like equation for the propagation of the photon-ALP beam through a magnetic field model. We briefly introduce the code here. \alpro\ solves the ALP-photon mixing problem numerically by formulating transfer matrices, following the method outlined by, e.g., \cite{de_angelis_relevance_2011}. The code is written in Python but uses just-in-time (JIT) compilation as part of the \textsc{numba} library \citep{numba} to speed up the matrix operations. \alpro\ is written in a modular fashion and includes routines for setting up various magnetic field models, including random turbulent fields and uniform field models, which will be expanded upon in future. The code is publicly available at \url{https://github.com/jhmatthews/alpro}, with documentation hosted on ReadTheDocs. We have tested the results from our code against analytic results as well as numerical results from the code used by \cite{marsh_new_2017} and the \textsc{gammaalps} code \citep{meyer_detecting_2014,meyer_gammaalps_2021}, finding excellent agreement.
\alpro\ was used for the ALP survival probability curves used by \cite{sisk_reynes_new_2021}, and the code includes the adaptive treatment of resonances described therein (see their Appendix B).
\alpro\ also contains an implementation of the Fourier formalism described by \cite{marsh_fourier_2022} and briefly outlined in section~\ref{sec:fourier}. This functionality allows the user to take advantage of fast Fourier transform (FFT) techniques to solve the ALP propagation problem in the massive and massless regime, accurate for relatively small conversion probabilities with amplitudes of a few per cent (see \citealt{marsh_fourier_2022} and section~\ref{sec:fourier_app} for more details on the valid regime for this formalism).

\section{The Magnetic Field in Perseus and Other Clusters}
\label{sec:icm}
For a given $m_a$ and $\g$, the quantities that determine the true $P_{\gamma\gamma}(E)$ are the perpendicular magnetic field along the line of sight to the X-ray point source, $B_\perp (z)$, and the plasma frequency and therefore electron density along the line of sight, $n_e(z)$. Thus, in addition to $n_e$, both the strength and structure of the field matter. We now review what is known about magnetic fields in clusters, focusing particularly on the well-studied Perseus cluster, and discuss the models for the magnetic field used in X-ray ALP searches to date. Throughout this section we will use $z$ as the radial coordinate for any spherically symmetric density or magnetic field profile, for consistency with the previous section. 

\subsection{Observational constraints on magnetic field strength and structure in the ICM}
Estimating the strength and structure of magnetic fields in the ICM is challenging. Constraints come from Faraday rotation measures (RMs), synchrotron-emitting radio haloes and relics, and the thermal pressure profile of the clusters \citep[see reviews by][]{carilli_cluster_2002,govoni_magnetic_2004}. We briefly review what is known about ICM magnetic fields from some of these approaches, with a particular focus on cool-core clusters such as Perseus. 

Arguably the most direct probes of ICM magnetic fields are Faraday RMs, which provide a measure of the parallel component of the magnetic field, $B_\parallel$, as defined by 
\begin{equation}
    {\rm RM} = 812~{\rm rad~m^{-2}} \int 
    \left(\frac{B_\parallel}{\mu{\rm G}}\right) 
    \left(\frac{n_e}{{\rm cm}^{-3}}\right) 
    \frac{dz}{{\rm kpc}} 
\label{eq:rm}
\end{equation}
where the integral is evaluated along the line of sight. An advantage of an RM measurement is it probes the same sightline that is traversed by the hypothetical photon-ALP beam; however, it gives no direct measure of $B_\perp$ and one also needs to know the electron density, $n_e$. Nonetheless, we still expect this method to give a reasonable handle on the integrated magnetic field along the line of sight, and at the very least it can be used as a check -- any field and density model adopted should not produce RMs that are, on average, dramatically in excess of that observed. Faraday RM maps against extended polarized sources can also be produced, which allow the coherence length and power spectrum of the magnetic field to be inferred \citep[e.g.][]{ensslin_magnetic_2003,murgia_magnetic_2004}. In general, Faraday RM observations of clusters are consistent with Kolmogorov turbulence, with magnetic field coherence lengths on the order of a few to tens of kpc \citep{feretti_magnetic_1995,feretti_radio_1999,allen_chandra_2001,clarke_new_2001,vogt_bayesian_2005,enslin_magnetic_2006,guidetti_intracluster_2008,govoni_rotation_2010,bonafede_coma_2010,kuchar_magnetic_2011}. The magnetic field strengths inferred are typically $B\sim~1\mu{\rm G}$, rising to $10$s of $\mu{\rm G}$ in the inner regions of cool-core clusters such as Hydra and Perseus. In the core of the Perseus cluster specifically, \cite{taylor_magnetic_2006} find substantial RMs in the range $6500-7500~{\rm rad~m^{-2}}$. The magnitude of these RMs are consistent with being produced by the ICM, but the pc-scale gradient in RMs reported by \cite{taylor_magnetic_2006} is harder to reconcile with the kpc-scale fields expected in the ICM, and a contribution from compact, dense filaments may be needed. 

Energetic constraints on ICM magnetic field strengths can be derived in synchrotron radio haloes and mini-haloes as well as from gamma-ray observations. Synchrotron radio haloes typically provide field estimates of a few $\,\mu {\rm G}$ under the assumption of minimum energy \citep[e.g.][]{giovannini_halo_1993,feretti_radio_1999,govoni_magnetic_2004,bonafede_giant_2014,kale_how_2016}, and these field strengths can easily be higher depending on the fractional pressure of non-radiating particles and the filling factor and geometry of the emitting material. Application of the hadronic minimum energy described by \cite{pfrommer_estimating_2004} to Perseus gives a (model-dependent) lower limit on the magnetic field of $4-9\,\mu {\rm G}$ in Perseus \citep{aleksic_constraining_2012}.  

The radial profile of the magnetic field strength can also be estimated by considering the pressure profile of the cluster. In this case, the thermal pressure as a function of radius can be estimated from deprojected X-ray observations under the assumption of spherical symmetry \citep[e.g.][]{russell_direct_2008}, or from the thermal Sunyaev-Zel`dovich effect \citep[e.g.][]{planck_collaboration_planck_2013}.
A magnetic field strength can then be calculated by assuming a plasma $\beta$, which we denote $\betapl$, defined (in Gaussian units) as 
\begin{equation}
    \betapl = \frac{P_{\rm th}}{\left( B^2 / 4 \pi \right)},
\end{equation}
where $P_{\rm th}$ is the thermal pressure. Although the value of $\betapl$ is uncertain, the canonical value is $\betapl \sim 100$ \citep[e.g.][]{bohringer_cosmic_2016,donnert_magnetic_2018}, which can be estimated from comparison of the inferred magnetic pressures from RM studies with the observed thermal pressures. $\betapl \sim 100$ is also expected if the magnetic energy density is in rough equipartition with the kinetic energy, since the velocity fluctuations in clusters and Perseus specifically are observed to be around $10-20$ per cent of the sound speed \citep[e.g.][]{zhuravleva_turbulent_2014,hitomi_collaboration_hitomi_2018}. Further discussion of the value of $\betapl$ in clusters in the context of ALP studies is given by \cite{marsh_fourier_2022} and \cite{sisk_reynes_new_2021}. For a constant $\betapl$, the sensitivity of the inferred limits on $\g$ to the value of $\betapl$ is straightforward. The transformation $B_\perp \rightarrow f B_\perp$ and $\g \rightarrow \g/f$, where $f$ is a constant, leaves the conversion probability unchanged. Thus, as long as $\betapl$ is uniform in $z$, increasing $\betapl$ by a factor $f^2$ translates into a weaker limit on $\g$ by a factor $f$. For a $\betapl$ that varies with $z$, the effect is more nuanced and depends on the relative importance of different regions of the cluster in determining the ALP signal. An example of how a variable $\betapl$ with radius -- scaling as $\betapl(z) \propto \sqrt{z}$ -- would affect the magnetic field profile in Perseus is shown in Fig.~\ref{fig:pressure_Bfield}; we examine the impact on the limits obtained in section~\ref{sec:limits}.

Finally, it is important to consider the physics of turbulence in clusters from both an observational and theoretical perspective. The Perseus cluster is clearly a dynamic, disturbed environment, with X-ray images showing bubbles, ripples and variations in surface brightness, particularly in the cluster core \citep[e.g][]{fabian_chandra_2000,fabian_very_2006,sanders_non-thermal_2005,sanders_deeper_2007,zhuravleva_turbulent_2014,walker_split_2018}. The amplitude and power spectrum of velocity fluctuations can be estimated from density fluctuations inferred from observed surface brightness fluctuations \citep{zhuravleva_turbulent_2014}, giving results that are consistent with Kolmogorov-like turbulence and supporting the emerging picture from ICM Faraday RMs described above. A turbulent ICM is not surprising -- while the physics depends on the details of the viscosity, the ICM is thought to have a moderately high effective Reynolds number \citep{schekochihin_plasma_2005,donnert_magnetic_2018}, meaning that `stirring' of the cluster on large-scales will transfer energy to small-scale turbulence via a Kolmogorov-like cascade. The driving scale of the turbulence can be crudely estimated from the Ozmidov scale, the scale on which the turbulent eddy turnover timescale becomes shorter than the buoyancy timescale. \cite{zhuravleva_turbulent_2014} estimate the Ozmidov scale at $\sim10$s of kpc in the Perseus cluster. Work to understand how the turbulent kinetic energy of the ICM is transferred to magnetic fields is ongoing, but important processes include the small-scale fluctuation dynamos \citep{schekochihin2004,schekochihin_plasma_2005}, as well as buoyancy and magnetothermal instabilities \citep{balbus2000,balbus2001,balbus2010,perrone_2021}.

Overall, observations imply the presence of $\sim 10\mu{\rm G}$ strength ICM magnetic fields in cool-core clusters and Perseus specifically. There are also good theoretical and observational reasons to expect the ICM magnetic field to be turbulent on $\sim$kpc scales. We will proceed by using these constraints to design appropriate magnetic field models assuming that a turbulent magnetic field is well-motivated; however, we will also investigate large-scale, `regular' magnetic fields by using a stochastic model that is coherent on $\gtrsim 50\,{\rm kpc}$ scales. 

\subsection{Models used in photon-ALP searches to date}
\begin{figure}
    \centering
    \includegraphics[width=\linewidth]{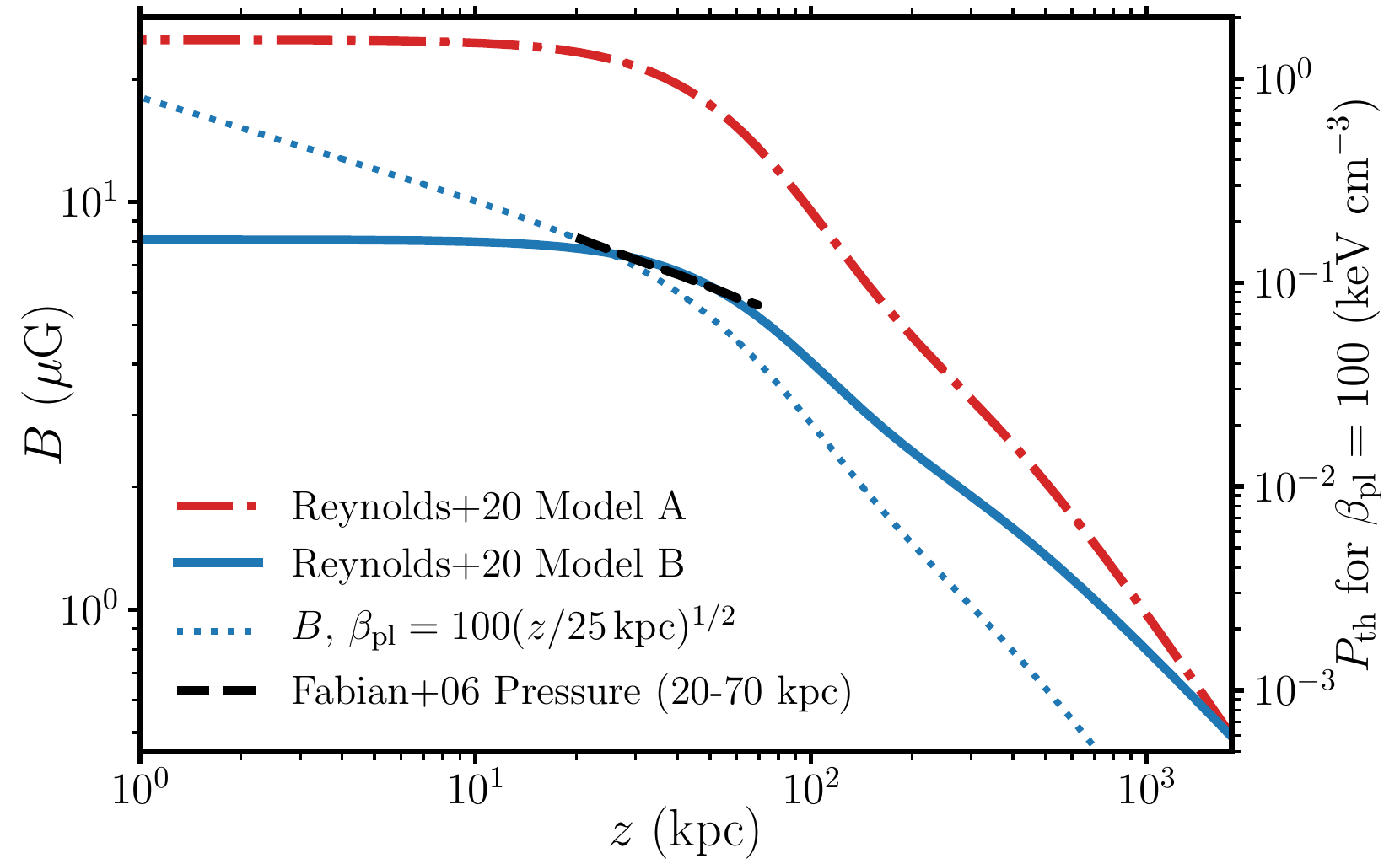}
    \caption{
     Magnetic field strength as a function of radius for Models A and B from \citetalias{reynolds_astrophysical_2020}, compared to the magnetic field inferred from the power-law pressure profile between 20 and 70 kpc from \citet{fabian_very_2006}, assuming $\betapl = 100$. The dotted line shows the magnetic field strength adjusted from Model B using a variable $\betapl(z) = 100(z/25\,{\rm kpc})^{1/2}$. The right-hand $y$-axis shows the corresponding thermal pressure for $\betapl = 100$ for Models A and B. By design, Model B matches the thermal pressure at $\approx25$\,kpc for $\betapl=100$, but Model A necessitates a significantly lower value of $\betapl$ to avoid over-predicting the thermal pressure at this radius. The variable $\betapl$ model, which only uses the left-hand $y$-axis, has a slightly stronger field compared to Model B in the inner $10~{\rm kpc}$ but a significantly weaker field on large scales.
    }
    \label{fig:pressure_Bfield}
\end{figure}

A common way of parameterising the magnetic field strength as a function of distance from the cluster centre, $z$, is using a power-law function of density such that 
\begin{equation}
    B(z) = B_0 \left[ \frac{n_e(z)}{n(R_0)} \right]^{\alpha},
    \label{eq:b_profile}
\end{equation}
where $\alpha$ is an exponent typically in the range $0-1$ and $R_0$ is some scaling radius, with $B_0 \equiv B(R_0)$ providing the normalisation of the field. To model $B(z)$ in Perseus, \citetalias{reynolds_astrophysical_2020} used two different models for the magnetic field, motivated by previous studies. For Model A, they adopted $R_0=0$, $B_0=25\,\mu {\rm G}$ and $\alpha=0.7$. Model A is a slightly altered version of the model used by \cite{berg_constraints_2017}, and is based on very long baseline array (VLBA) observations of NGC 1275. 
For Model B, they used $R_0=25\,{\rm kpc}$, $B_0=7.5\,\mu {\rm G}$ and $\alpha=0.5$. This model is instead based on the radial pressure profile derived by \cite{fabian_very_2006} from deep X-ray observations (with a total exposure time of $900$\,ks) of the Perseus cluster, assuming $\betapl = 100$. We show the magnetic field strength $B(z)$ for Models A and B of \citetalias{reynolds_astrophysical_2020} in Fig.~\ref{fig:pressure_Bfield}. The thermal pressure profile, $P_{\rm th}$ calculated with $\betapl = 100$, is plotted on a twin $y$-axis. We also show the magnetic field strength inferred from the power-law approximation to the thermal pressure between $20$ and $70$\,kpc from \cite{fabian_very_2006}. Model B is based on this pressure profile, so agrees well with the observed data, but Model A substantially over-predicts the thermal pressure due to a more optimistic estimate for the magnetic field strength throughout the cluster volume; $\betapl \sim 10$ would be needed to bring agreement with the observed $P_{\rm th}$, which is lower than expected. We therefore adopt the more realistic form of $B(z)$ from  \citetalias{reynolds_astrophysical_2020} model B for our work (but we also consider a model with variable $\betapl(z)$ in subsequent sections). 

It is the perpendicular component of $\myvector{B}(z)$ that appears in the ALP mixing matrix (equations~\ref{eq:mixing}-\ref{eq:delta_y}), so the direction of the field matters, as does its coherence and isotropy. One possible way to model turbulent magnetic fields is using a Gaussian random field (GRF), in which a random, isotropic field is generated according to some power spectrum and then `shaped' so that the field strength decays with radius. GRF models have been used in ALP cluster studies in a number of cases \citep{wouters_axion-like_2012,angus_soft_2014,meyer_detecting_2014,schallmoser_updated_2021}. Alternatively, a `cell-based' model can be used, in which the magnetic field along the line of sight is modelled as a series of cells of extent $\Delta z$ \citep[e.g.][]{wouters_axion-like_2012}. Each cell is an approximation to a given patch of turbulent field and therefore has a size comparable to the coherence length of the field, $\Lambda_c$, a random and isotropic field direction in each cell. $\Delta z$ is chosen according to a probability distribution function $p(\Delta z)$, typically a power-law. For example, \citetalias{reynolds_astrophysical_2020} use $p(\Delta z) \propto \Delta z^{-1.2}$ spanning $3.5-10$\, kpc. In their model B, \citetalias{reynolds_astrophysical_2020} scale these minimum and maximum cell sizes linearly with radius as $z/50{\rm kpc}$, because coherence lengths are expected to grow with distance from the cluster centre. 

The cell-based method has been used in X-ray studies of M87 \citep{marsh_new_2017}, NGC 1275 \cite{reynolds_astrophysical_2020} and H1821+643 \citep{sisk_reynes_new_2021}, and is also discussed from a theoretical and modelling perspective by \cite{{wouters_axion-like_2012}}, \cite{galanti_behavior_2018} and \cite{marsh_fourier_2022}. Indeed, \cite{galanti_behavior_2018} suggest that the discontinuities inherent to the cell-based method can introduce unphysical results in the conversion probability. Cell-based models have continuous magnetic autocorrelation functions consisting of piece-wise linear segments, joined at kinks. In the perturbative regime, structure in this autocorrelation function maps directly to structure in the conversion probability that can be interpreted using Fourier analysis \citep{marsh_fourier_2022}. Specifically, in this regime, the conversion probability from a cell-based model (and, in fact, any discretised model) can be understood as a combination of two effects: (i) an incoherent sum of the oscillatory pattern introduced by each cell and (ii) interference terms whose frequency of oscillation in $1/E$ is set by the cell boundaries. The jaggedness of the autocorrelation functions of cell-based models leads to enhanced support at high conjugate frequencies, and in the perturbative formalism, this can be understood as excess conversion probability at low energies \citep{marsh_fourier_2022}. In many cases of interest, this artificial feature of the cell-models only affects the conversion probability below the energy range where it is maximised, and is hence of limited observational importance. For cell models with a constant cell size $\Delta z$, unphysical features can be produced with a reciprocal energy spacing that depends on $\Delta z$; in the massive ALP regime, this spacing is $\Delta (1/E) = 2 \pi / (m_a^2 \Delta z)$ where $\Delta z$ is in natural (eV) units. However, in practice, a distribution for $\Delta z$ is often used and $\Delta z$ is usually fairly small compared to the total path length; both these factors act to wash out and decrease the power in this class of artificial features. Nevertheless, cell-based models are still clearly an approximation to a complex, continuous magnetic field structure. 

In addition to $\myvector{B}(z)$, the other important cluster quantity for photon-ALP mixing is the electron density $n_e$ which determines the ICM plasma frequency through $\omegap=\sqrt{4\pi n_e e^/m_e}$. The density in clusters decreases with radius and is often modelled using a so-called $\beta$--law, given by 
\begin{equation}
n_e(z) = \left[ 1 - \left(\frac{z}{r_c}\right)^2 \right]^{-3\beta/2}
\label{eq:betalaw}
\end{equation}
where $r_c$ is a core radius and $\beta$ an exponent of order unity. \citetalias{reynolds_astrophysical_2020} instead set the density using a double $\beta$--law, proposed by \cite{churazov_xmm-newton_2003} as an analytic approximation to the electron density in the Perseus cluster
\begin{equation}
n_e(z) = \frac{3.9\times10^{-2}}{\left[1+ (z/80\, {\rm kpc})^2\right]^{1.8}} + \frac{4.05\times10^{-3}}{\left[1+ (z/280\, {\rm kpc})^2\right]^{0.87}}
~{\rm cm}^{-3}.
\label{eq:ne}
\end{equation}
We do expect the electron density to have an impact on the inferred limits on ALP parameters, especially in setting the constraints on $\g$ at the high mass end. However, given that the density is usually well determined from X-ray observations we have chosen to focus on the impact of the magnetic field, rather than the density profile, in our work, so we proceed in using equation~\ref{eq:ne} for all our photon-ALP conversion calculations hereafter. 

\begin{figure*}
\centering
\includegraphics[width=0.9\linewidth]{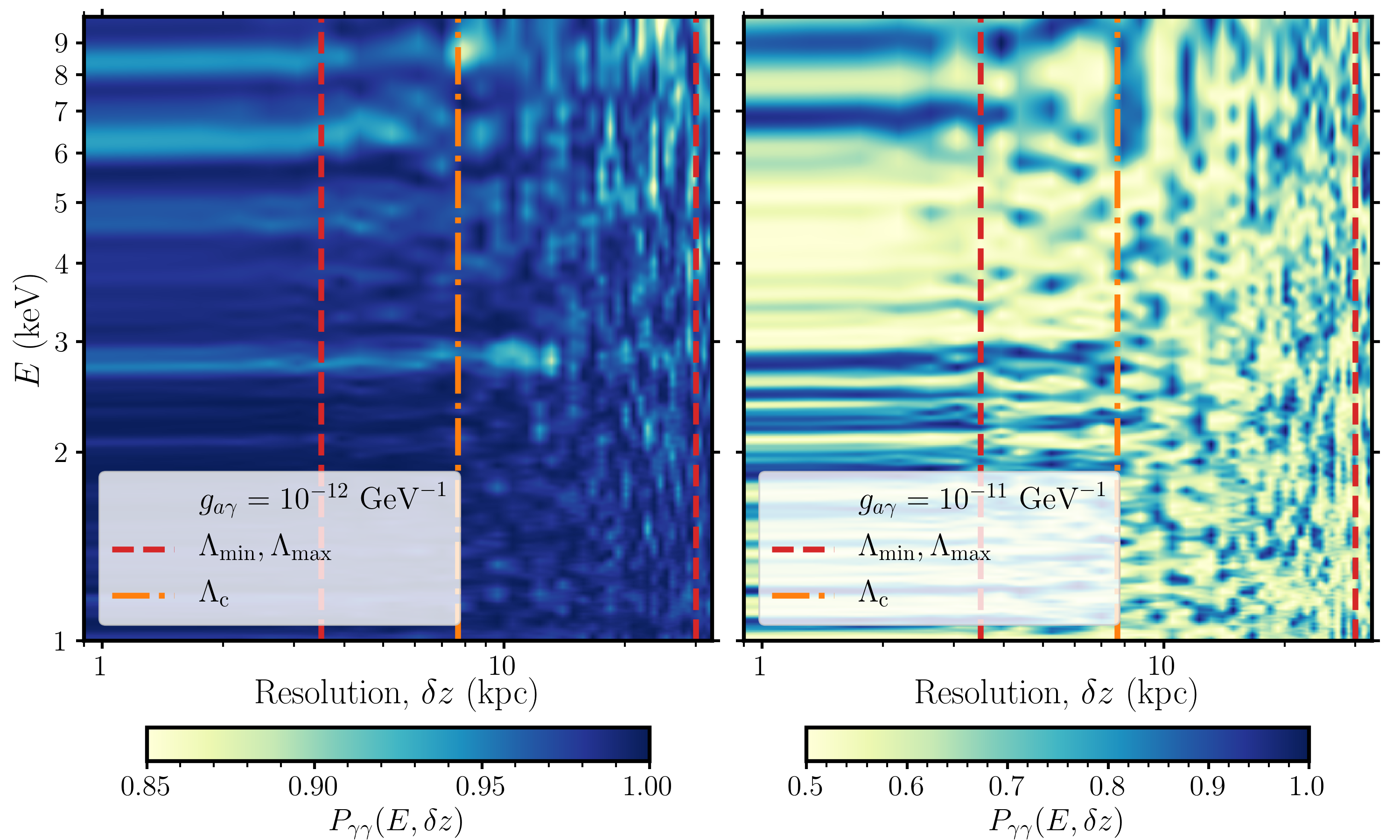}
\caption{The impact of resolution on a single ALP survival curve. The conversion probability is shown as a function of energy $E$ and resolution $\dz$ for a single GRF model realization, for two different values of the coupling constant ($\g=10^{-12}$\invgev, left, and $\g=10^{-11}$\invgev, right), and for $m_a = 10^{-13}$\,eV. Vertical red dashed lines mark the minimum and maximum scale lengths of the power spectrum ($\lmin$ and $\lmax$) and the orange dot-dashed line marks the coherence length $\lc$. The probability stops changing significantly around $\lmin$, showing that structure below the coherence length affects the curves and it is important to adequately resolve the field model.}
\label{fig:pgg_seed0}
\end{figure*}

\section{ALP-photon conversion in turbulent fields: Model sensitivity}
\label{sec:sensitivity}
As discussed above, there are a number of different ways of modelling a turbulent cluster magnetic field, with the cell-based and GRF methods being the most common. We will first examine the influence of small-scale field structure using a divergence-free GRF, before discussing the ALP survival probability from a mixture of cell-based and GRF models with varying approximations. A turbulent magnetic field can be characterised by its power spectrum, which generally takes a power-law form, such that $E_k dk \propto k^{-n} dk$, where $n$ is a power-law index, $k=2\pi/\Lambda$ is the wavenumber for a given wavelength $\Lambda$, and $E_k dk$ is the energy contained in the interval $(k,k+dk)$. For Kolmogorov turbulence, $n=5/3$. The steeper the index, the more energy is contained (in relative terms) on large scales. It is useful to define the coherence or correlation length of the field, which is the approximate scale upon which the field becomes decorrelated from a neighbouring `patch'. For power-law, isotropic turbulence the coherence length is given by \citep[e.g.][]{harari_lensing_2002}
\begin{equation}
    \Lambda_c = \frac{\lmax}{2} \frac{n-1}{n} \frac{1 - (\lmax/\lmin)^n}{1 - (\lmax/\lmin)^{n-1}} \, ,
\end{equation}
where $\lmin$ and $\lmax$ are the minimum and maximum scale lengths of the turbulence (over which the power spectrum is defined). By considering a few instructive limits we can see how $\Lambda_c$ changes for different dynamic ranges and power spectrum indices.  In the case of turbulence with a large dynamic range and Kolmogorov index $n=5/3$, $\Lambda_c \rightarrow 0.2 \Lambda_{\rm max}$ as $\Lambda_{\rm max}/\Lambda_{\rm min}\rightarrow \infty$. For very sharply peaked turbulence ($n \gg 1$ or $\Lambda_{\rm min}\rightarrow \Lambda_{\rm max}$), $\Lambda_c \rightarrow 0.5 \Lambda_{\rm max}$. 
Thus, in general, the coherence length can be significantly larger than the smallest scale length. As a result, there can be significant structure and energy contained in modes with $\Lambda<\Lambda_c$. 

\subsection{Generating Random Gaussian Fields}
To generate random Gaussian magnetic fields, we follow the approach described by \cite{tribble_radio_1991}, and discussed further by various authors \citep{murgia_magnetic_2004,hardcastle_synchrotron_2013,angus_soft_2014}. We first generate random Fourier amplitudes drawn from a Rayleigh distribution, such that the probability density of amplitude $\Acal$ is
\begin{equation}
    P(\Acal,\varphi) d\Acal d\varphi = \frac{\Acal}{2 \pi \Acal_k^2} \exp \left(-\frac{\Acal^2}{\Acal_k^2} \right) d\Acal d\varphi.
\end{equation}
where the amplitude is a power-law of the form $\Acal_k \propto k^{-\zeta}$ defined between minimum and maximum wavenumbers $k_{\rm min} = 2\pi / \Lambda_{\rm max}$ and $k_{\rm max} = 2\pi / \Lambda_{\rm min}$. The index $\zeta$ is related to $n$ by $\zeta = n + 2$. We draw phases $\varphi$ uniformly between 0 and $2\pi$. We then take the inverse Fourier transform of the Fourier amplitudes and phases to obtain $\myvector{A}$, the vector potential in real space. We apply the magnetic field radial profile (e.g. from equation~\ref{eq:b_profile}) before calculating the magnetic field as $\myvector{B} = \nabla \times \myvector{A}$, resulting in a (numerically) divergence-free magnetic field ($\nabla \cdot \myvector{B} = 0$). We generate a 3D GRF model and then take a single 1D sightline to the centre as the input to our photon-ALP conversion calculation. The parameters describing a GRF model are $\zeta$, the minimum and maximum scales $\Lambda_{\rm min}$ and $\Lambda_{\rm max}$, and the domain size $z_{\rm max}$. For a given field model, the calculation of the ALP survival probability then uses $N_z$ cells or Fourier samples, which determines the spatial resolution of the model, $\dz$.

\subsection{Sensitivity to small scale field structure}
\label{sec:smallscale}
To examine the sensitivity to small-scale field structure for strong ALP signals, we conducted a resolution test. We first generated 64 different realizations of a GRF model using the procedure described above, with model parameters $\lmin=3.5$\,kpc, $\lmax=30$\,kpc, $n=5/3$ and $z_{\rm max}=1.8$\,Mpc. The profile $B(z)$ used to shape the magnetic field is that of \citetalias{reynolds_astrophysical_2020}'s Model B, from equation~\ref{eq:b_profile} with $R_0=25$\,kpc, $\alpha=0.5$ and $B_0=7.5\,\mu{\rm G}$. Together, this choice of parameters corresponds to model 4 as described in section~\ref{sec:field_models}, and leads to a coherence length of $\lc=7.67$\,kpc. We then calculated survival probability curves in the $1-10$\,keV range at a range of spatial resolutions $\dz$, sampling in the range $0.25\lmin \leq \dz \leq 2.5\lmin$ at $0.25 \lmin$ intervals and the range $2.5\lmin \leq \dz < 20\lmin$ at $0.5 \lmin$ intervals. We focus on the low mass ALP case with $m_a = 10^{-13}$~eV, and calculate curves at different coupling constants. We first consider a single field realization and plot the survival probability $\pgg$ as a function of $\dz$ and energy $E$ in Fig.~\ref{fig:pgg_seed0}, for $\g=10^{-12},10^{-11}$~GeV$^{-1}$. We do not present results for lower values of $\g$ because they are almost identical to the $\g=10^{-12}$\invgev\ case, but with $\pga$ scaled by a factor of $\g^2$. The survival probability stops changing significantly around $\lmin$, showing that structure below the coherence length can be important. It is therefore necessary to resolve the minimum scale length of the magnetic field in these cases to get an accurate survival probability. 

\begin{figure}
\includegraphics[width=\linewidth]{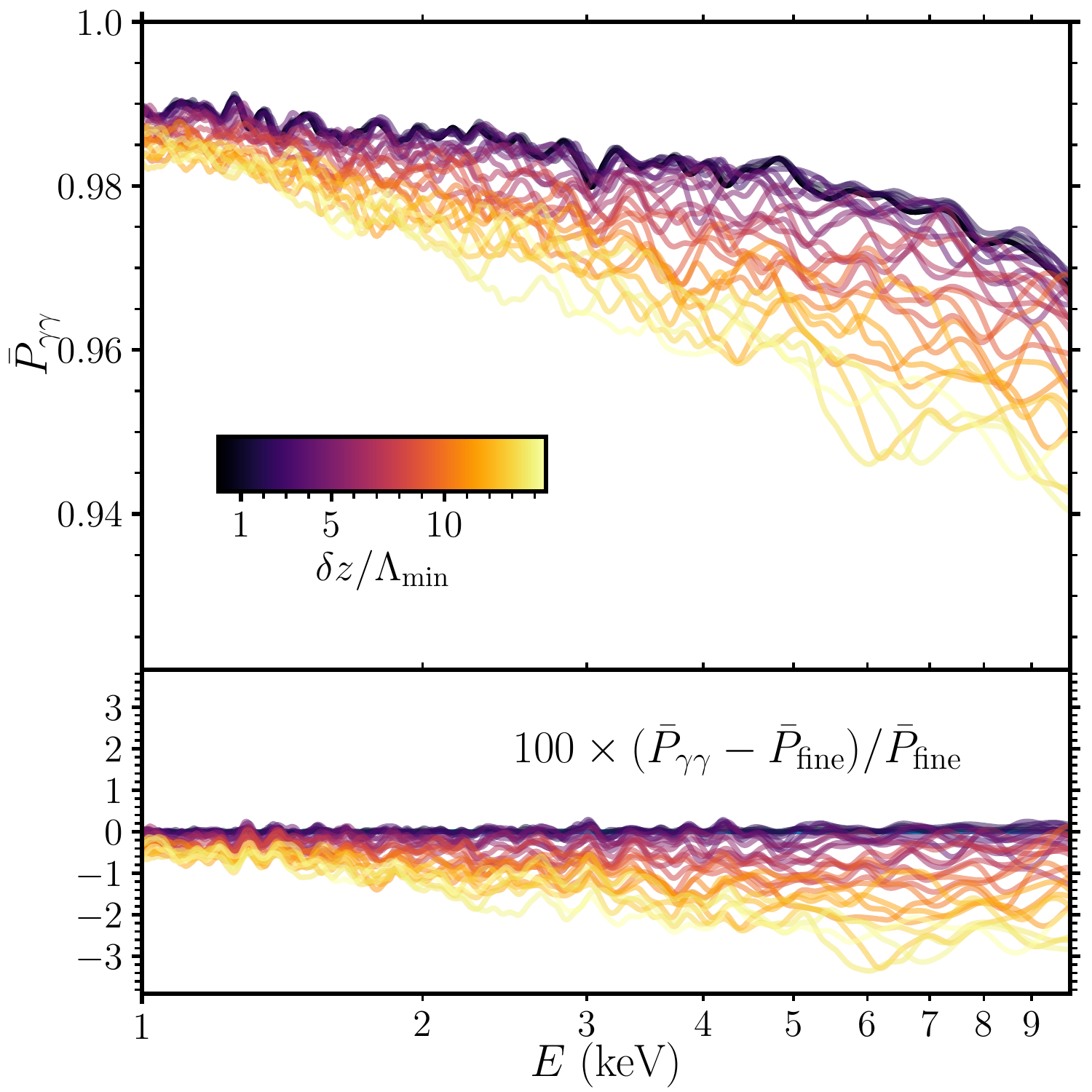}
\caption{
Mean survival probability for $m_a=10^{-13}$\,eV and $\g=10^{-12}$\,\invgev, colour-coded by resolution $\dz$, calculated from $N=64$ realisations of the GRF model described in section~\ref{sec:smallscale} (model 4 in latter sections). The bottom panel shows the percentage residual with respect to $\bar{P}_{\rm fine}$, defined as the mean survival probability of the model at the finest resolution (i.e. the closest to the true survival probability).}
\label{fig:mean_res}
\end{figure}

We have checked this result holds for different field realisations, but to show this explicitly we can consider the mean survival probability at each energy marginalised over random number seed, given by $\bar{P}_{\gamma \gamma}(E) = [\sum_i \pgg(E,i)]/N$. This quantity is plotted in Fig.~\ref{fig:mean_res} as a function of energy and colour-coded by resolution $\dz$ for $N=64$, $\g=10^{-12}$~GeV$^{-1}$ and $m_a = 10^{-13}$~eV. In the bottom panel, we show the (percentage) residual compared to the mean survival probability at the finest resolution ($\dz \approx \lmin/4$), $\bar{P}_{\rm fine}$, which is assumed to correspond to `ground truth' for this type of field model. Once again, we see that the survival probability converges around $\lmin$. Models that are significantly under-resolved can under-predict the survival probability (and thus over-predict the impact of ALPs) by a few per cent. 

\subsection{Comparison of ALP signals from five different magnetic field models}
\label{sec:field_models}

Our aim is to investigate how the photon-ALP survival probabilities, and resulting 
limits on photon-ALP coupling, $\g$, depend on the magnetic field model used. To this end, we use five different field models:
\begin{itemize}
    \item Model 1: A cell-based magnetic field model as Model B in \citetalias{reynolds_astrophysical_2020}. 
    \item Model 2: As model 1, but without the linear scaling of cell-size $\Delta z$ with radius. 
    \item Model 3: As model 1, but with a variable $\betapl (z)$ such that $B(z)$ is scaled by $[100/\betapl (z)]^{1/2}$. This model is designed to be consistent with the available magnetic field constraints and $\betapl=100$ in the cluster core, but allows for the $\betapl (z)$ to increase with distance from the cluster centre. Such an effect could be produced if, for example, magnetic field amplification is less effective in the outer regions of the cluster. This choice of $\betapl (z)$ is more conservative in terms of the value of $B_\perp$ in the cluster outskirts. 
    \item Model 4: A Gaussian random field model with Kolmogorov power spectrum and minimum and maximum scale lengths of 3.5 kpc and 30 kpc, respectively.
    \item Model 5: A Gaussian random field model with Kolmogorov power spectrum and minimum and maximum scale lengths of $25.125$\,kpc and $225$\,kpc, respectively. This model is designed to approximate magnetic fields that are coherent on fairly large-scales in the cluster core, but still allows us to marginalise over multiple field realisations.
\end{itemize}
Parameters and more details on the models are given in Table~\ref{table:models}. We adopt a minimum radius of $10$\,kpc for all our calculations, which is slightly more conservative than \citetalias{reynolds_astrophysical_2020}, and a maximum radius of $1.8$\,Mpc, the virial radius of the Perseus cluster (see \citealt{sisk_reynes_new_2021} for a discussion of the sensitivity to these parameters for the H1821$+$643 limits). The GRF models use a resolution of $\dz = \lmin$ informed by the sensitivity study in the previous subsection. Although slightly {\sl ad hoc}, the choices of scale lengths for models 4 and 5 are made to mimic ICM magnetic fields with qualitatively different structures: model 4 as an approximation to the kpc-scale turbulence typically observed in cool-core clusters, and model 5 to imitate larger scale coherent modes, which may, for example, be produced by AGN activity. The limitations of these models are discussed further in section~\ref{sec:other_limits}.

\begin{figure}
    \centering
    \includegraphics[width=\linewidth]{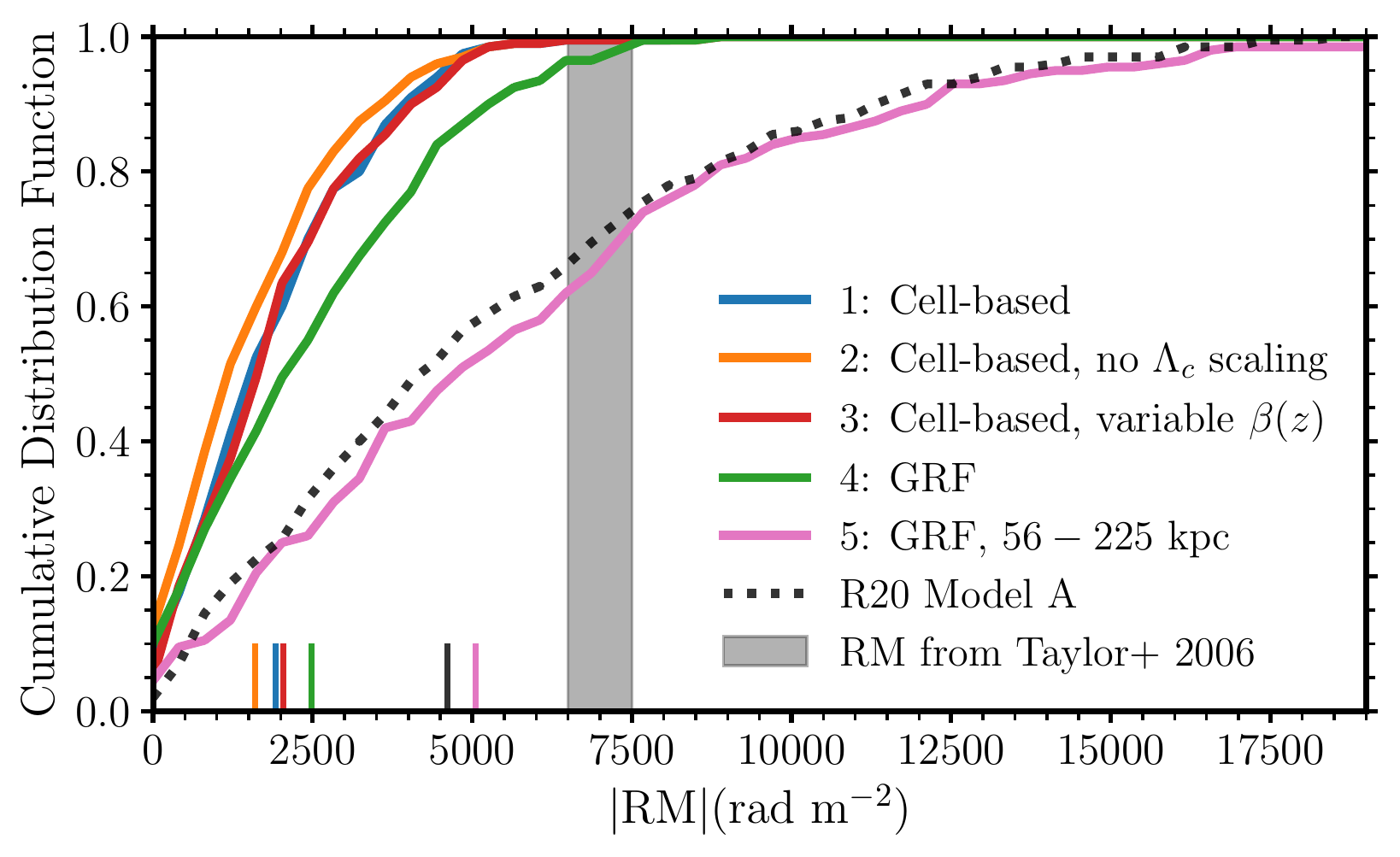}
    \caption{
    Cumulative distribution function of the absolute RMs from 200 realisations of the five magnetic field models used in this work, described in section~\ref{sec:field_models}. We also show the RMs from 200 realisations of model A from \citetalias{reynolds_astrophysical_2020}. The median absolute RM for each model is marked with a vertical line on the $x$-axis. The shaded region shows the range of RMs reported by \citet{taylor_magnetic_2006} from observations of the Perseus core ($6500-7500\,{\rm rad~m^{-2}}$).
    }
    \label{fig:rm_cdf}
\end{figure}

To examine the Faraday RMs predicted by this set of models, we show the cumulative distribution function (CDF) of the absolute value of the RM in Fig.~\ref{fig:rm_cdf}, compared to the range of values from \cite{taylor_magnetic_2006}. Models 1-4 are quite conservative in their predictions of Faraday RMs, with $\geq 95$ per cent of the realisations producing $|{\rm RM}|$ below the lower bound of the \cite{taylor_magnetic_2006} measurement ($6500\,{\rm rad~m^{-2}}$). The median values of $|{\rm RM}|$ for these models lie in the range $\approx 1500-2500\,{\rm rad~m^{-2}}$ (see Table~\ref{table:models} for the exact values). Model 5 predicts slightly higher magnitude RMs, comparable to model A from \citetalias{reynolds_astrophysical_2020}, which is expected since more coherent fields with significant radial components produce higher RMs \citep{feretti_magnetic_1995}. However, the median $|{\rm RM}|$ from model 5 is still lower than the observed value, so this field prescription is still broadly consistent with observations. In fact, the expected $|{\rm RM}|$ measurement from models 1 to 5 is below the range inferred by \cite{taylor_magnetic_2006}, implying that our models make fairly reasonable and conservative predictions that are generally consistent with observations, albeit somewhat dependent on the exact nature of the Faraday screen. 

\begin{table*}
\centering 
\renewcommand{\arraystretch}{1}
\begin{tabular}{c c c c c c c c c}
%     5 & 200 & Partially covering absorber $+$ $6.4$~keV line &  GRF & No & constant, 100 & Yes & \color{C4}{$\blacksquare$}  \\
\hline
    Model & $N$ & $B$-Field & $\lc$ scaling? & $\betapl(z)$ & Range of scales (kpc) & $\lc$ (kpc) & Median $|{\rm RM}|\,({\rm rad\,m^{-2}})$ & Colour \\\hline \hline 
    1 & 200 &  Cell-based & Yes & constant, 100 & $3.5-10$ & -- & 1915 & \color{C0}{$\blacksquare$} \\
    2 & 200 &  Cell-based & No & constant, 100 & $3.5-10$ & -- & 1603 & \color{C1}{$\blacksquare$} \\
    3 & 200 &  Cell-based & Yes & $100~\sqrt{z/25\,{\rm kpc}}$ & $3.5-10$ & -- & 2045 & \color{C3}{$\blacksquare$}  \\
    4 & 200 &  GRF & No & constant, 100 & $3.5-30$ & $7.67$ & 2480 & \color{C2}{$\blacksquare$}  \\
    5 & 200 &  GRF & No & constant, 100 & $25.125-225$ & $57.1$ & 5050 & \color{C6}{$\blacksquare$}  \\
    \hline
\hline
\end{tabular}
\caption{
Magnetic field models used in this work in the calculation of photon-ALP survival probabilities (e.g. Fig.~\ref{fig:Bfield_and_pgg}) and to obtain limits on ALP parameters in Fig.~\ref{fig:lims_comparison}. The colours shown in the last column match the colours used in the relevant figures. Each model uses the same radial profile for $B(z)$ as model B from \citetalias{reynolds_astrophysical_2020}, except for model 3 which adjusts this by a factor $[100/\betapl(z)]^{1/2}$. 
}
\label{table:models}
\end{table*}

\begin{figure*}
    \centering
    \includegraphics[width=0.9\linewidth]{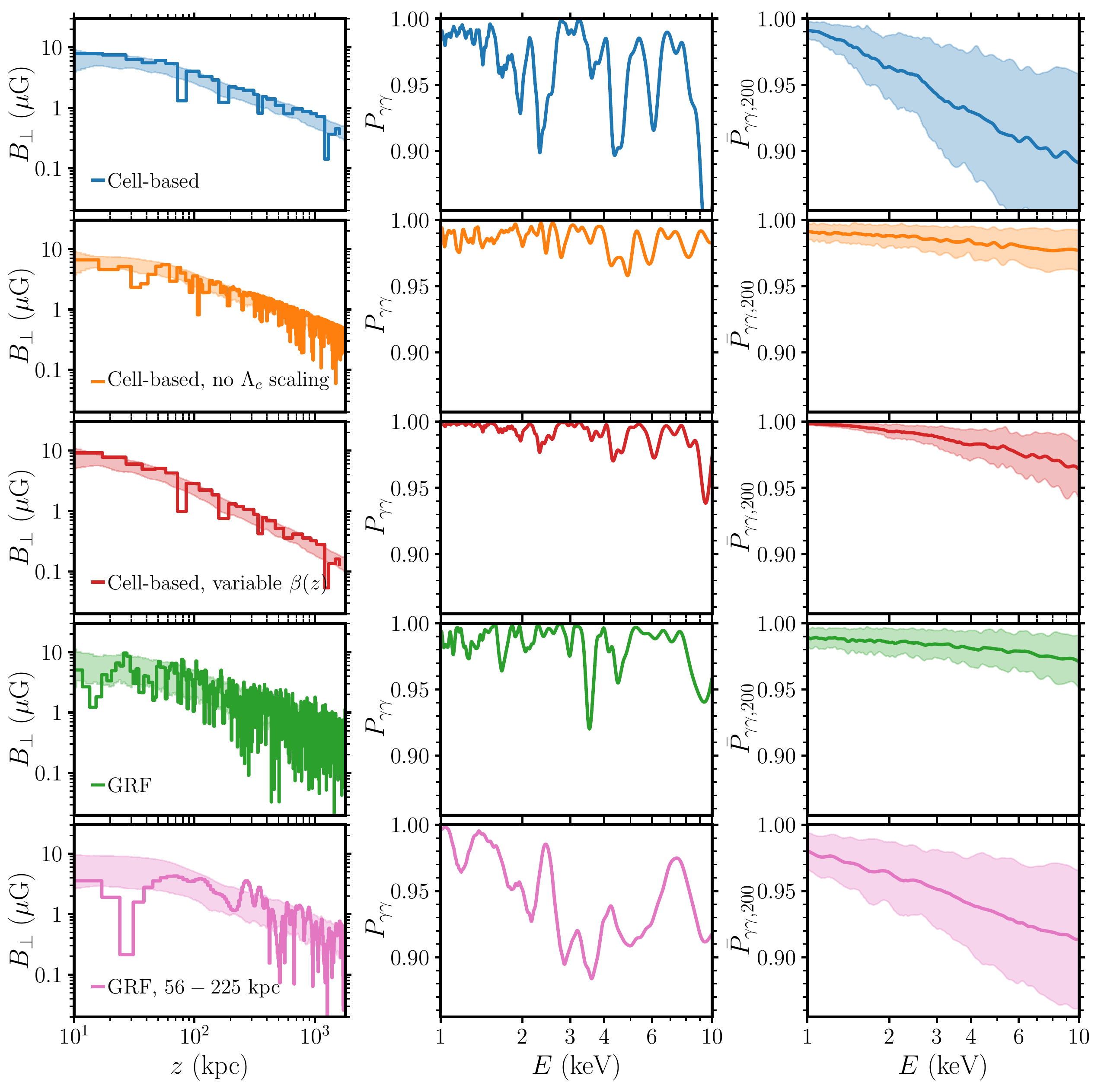}
    \caption{
    The magnetic field models used in this work and their associated photon-ALP survival probabilities. The colours corresponding to each model match those in Table~\ref{table:models} and Fig.~\ref{fig:lims_comparison}. {\sl Left:} the profile of the perpendicular magnetic field component, $B_\perp(z)$ for a single realisation of each field model, with the shaded region showing the root-mean-square range of all 200 models. {\sl Middle:} The survival probability $\pgg$ from the individual field realisation as a function of energy in the X-ray band. {\sl Right:} The mean survival probability averaged over 200 realisations, $\bar{P}_{\gamma\gamma,200}$, with the standard deviation of $\pgg$ in each energy bin shown as a shaded region. 
    }
    \label{fig:Bfield_and_pgg}
\end{figure*}

The line-of-sight profile of $B_\perp(z)$ from a single realisation of each of these models is shown in the left-hand panel of Fig.~\ref{fig:Bfield_and_pgg}. The shaded region shows the root-mean-square range of all 200 models. These $B_\perp(z)$ profiles are all stochastically generated using different approaches so there is no direct equivalence between the realisations, but their general characteristics can still be compared. Some of these characteristics are fairly trivial -- for example, we can see that model 3, with the variable $\betapl(z)$, has a decreased field strength at large radii compared to the fiducial model 1. Similarly, model 2 does not have the scaling factor applied to $\Delta z$, resulting in cell sizes that are uniform, rather than uniform in log-space like models 1 and 3. The cell-based models are clearly qualitatively different to the GRF model with small-scale structure (model 4). Firstly, the cell-based approach does not produce the same small-scale structure as the GRF model, because it has a minimum cell size of $3.5$~kpc and cells are on average significantly larger than this. Secondly, the dynamic range is smaller, because the field strength in each domain is set from equation~\ref{eq:b_profile} with $R_0=25\,{\rm kpc}$, $B_0=7.5\,\mu {\rm G}$ and $\alpha=0.5$, and so any variation in $B_\perp(z)$ relative to this results only from the random, isotropic choice of the direction of the vector. In contrast, although $B_\perp(z)$ averages to similar values in the GRF approach, all Fourier modes are accounted for and in some cases the field is significantly lower or higher than the domains approach, which translates into a larger range in $B_\perp(z)$; this effect is discussed by \cite{schallmoser_updated_2021}. 

\subsubsection{Survival Probabilities}
For each of our magnetic field models, we use \alpro\ to calculate photon survival probabilities $\pgg(E)$ at an energy resolution much finer than the data, using $200$ field realisations (as set by the random number seed) for each of the five magnetic field models. These ALP survival probability curves are used for calculating limits on ALP parameters in the next section, but we can also examine the form of $\pgg(E)$ in each case. The middle panel of Fig.~\ref{fig:Bfield_and_pgg} shows the survival probability from the same single realisation shown in the left-panel, while the right-hand panel shows the mean survival probability calculated from 200 realisations, denoted $\bar{P}_{\gamma\gamma,200}(E)$, with the standard deviation shown as a shaded region. There are some notable differences between the characteristics of the curves and the amplitude of the ALP signal. Models 2 and 4 both produce notably larger $\pgg$ and $\bar{P}_{\gamma\gamma,200}(E)$ (smaller $\pga$), particularly at higher $E$, compared to the fiducial cell-based model. Model 3, with the variable $\betapl(z)$ also produces weaker ALP signals, notably in this case at both low and high $E$, due to its decreased magnetic field strength at large radii. The larger-scale GRF model produces features that are quite broad in energy width and an amplitude that is comparable to Model 1 and larger than Models 2-4. 

\subsection{Summary of this section and literature comparison}
\label{sec:literature}

A number of authors have either applied GRF models to ALP cluster studies, considered the drawbacks of cell-based models, or made explicit comparison between GRF and cell-based models \citep{wouters_axion-like_2012,wouters_constraints_2013,meyer_detecting_2014,angus_soft_2014,galanti_behavior_2018,schallmoser_updated_2021,marsh_fourier_2022}. \cite{wouters_axion-like_2012} originally used a cell-based model for their investigation, but also compared results with those from a Kolmogorov spectrum GRF, finding a broad agreement in the variance of the residuals in their synthetic data. Based on arguments given by \cite{mirizzi_constraining_2009}, \cite{wouters_constraints_2013} suggest that, in Kolmogorov turbulence, the root-mean-square intensity of the magnetic field varies as $\Lambda^{1/3}$, leading to an approximate scaling of the conversion probability as $\pga \sim \Lambda^{2/3}$. In this case we might expect that the small-scale magnetic field would not have a significant impact on photon-ALP conversion. However, the exponent is relatively weak, and will affect small scale-lengths orders of magnitude below $\lc$. Since $\lc \sim \lmax / 5$ for broadband Kolmogorov turbulence, scales below the coherence length can still have a significant impact, at the $\sim 5$ per cent level. This approximate scaling is in agreement with our findings and shows that it is important to conduct ALP calculations with a reasonable dynamic range of scales and resolution. This is perhaps an argument against the cell-based models, since these models have no real structure below $\sim \lc$. Having said this, the form of the survival probability from cell-based models and GRF models is actually rather similar, as found in previous studies \citep{wouters_axion-like_2012,meyer_detecting_2014,schallmoser_updated_2021,marsh_fourier_2022}, so while small-scale structure can be important, it is probably a sub-dominant effect compared to the systematic uncertainty on the radial profiles of $\betapl$ and $\lc$. 

Overall, our calculations show that the magnetic field model has an impact on the form of $\pgg(E)$, which is sensitive to the radial profile and coherence scale of the field. Na\"ively, since $\pga \propto \g^2$, we can anticipate that smaller conversion probabilities by a factor 2 would translate to weaker limits on $\g$ by $\sqrt{2}$. However, the exact change in the limits is $m_a$-dependent and partly dictated by the signal-to-noise in each energy bin, which is a function of the intrinsic source spectrum and the observatory/instrument configuration.  We therefore conduct a re-analysis of the NGC 1275 {\it Chandra} data in the next section using the models described here (but over a wide range of $m_a$ and $\g$).

\section{A Re-analysis of the NGC 1275 X-ray data}
\label{sec:reanalysis}
We now turn to observational data to test the sensitivity of the limits to the effects discussed, using the same five magnetic field models we described in the previous section.

\begin{table*}
\centering 
\renewcommand{\arraystretch}{1.2}
\begin{tabular}{c c c c c}
\hline
    \multicolumn{5}{c}{Free Parameters} \\ \hline \hline 
    Component & Parameter & Description & HEG & MEG \\
    \hline 
    \texttt{pow} & $A_X$ & Power-law normalisation & $8.92 ^{+0.67}_{-0.63} \times 10^{-3}$ & $9.45^{+0.62}_{-0.58}  \times 10^{-3}$ \\
    \texttt{pow} & $\Gamma_X$ & Photon index & $1.92^{+0.04}_{-0.04}$ & $1.94^{+0.03}_{-0.03}$ \\ 
    \texttt{tbpcf} & $N_H~({\rm cm}^{-2})$ & Column density & \multicolumn{2}{c}{$7.54 ^{+4.24}_{-2.38} \times 10^{22}$} \\
    \texttt{tbpcf} & $f_{\rm cov}$ & Covering factor & \multicolumn{2}{c}{$9.50^{+4.53}_{-4.70}  \times 10^{-2}$} \\
    \texttt{zgauss} & $A_{\rm line}~({\rm phot\, cm}^{-2}\,{\rm s}^{-1})$ & Line normalisation & \multicolumn{2}{c}{$4.06^{+2.46}_{-2.29} \times 10^{-6}$}
    \\ \hline
    \multicolumn{5}{c}{Frozen Parameters} \\ \hline\hline 
    Component & Parameter & Description & HEG & MEG \\ \hline 
    \texttt{tbabs} & $N_H~({\rm cm}^{-2})$ & Column Density (Galactic) & \multicolumn{2}{c}{$1.32\times10^{21}$} \\ \hline
     \multicolumn{5}{c}{Fit statistic (without ALPs)} \\ \hline \hline 
    & $C/{\rm dof}$ & \multicolumn{3}{c}{4857/4863} \\
\hline
\end{tabular}
\caption{Parameters used in the X-ray spectral modelling, given with the best-fit values and error estimates obtained from a spectral fit without ALPs present. Uncertainties quoted are $90$ per cent confidence intervals as calculated using the \texttt{error} command in \xspec. All parameter values are quoted at 3 significant figures with uncertainties given to the same absolute precision. Free parameters spanning both HEG and MEG columns have their values tied.}
\label{table:fiducial}
\end{table*}

\defcitealias{reynolds_probing_2021}{R21}
\subsection{Observational data and spectral modelling}
\begin{figure}
    \centering
    \includegraphics[width=\linewidth]{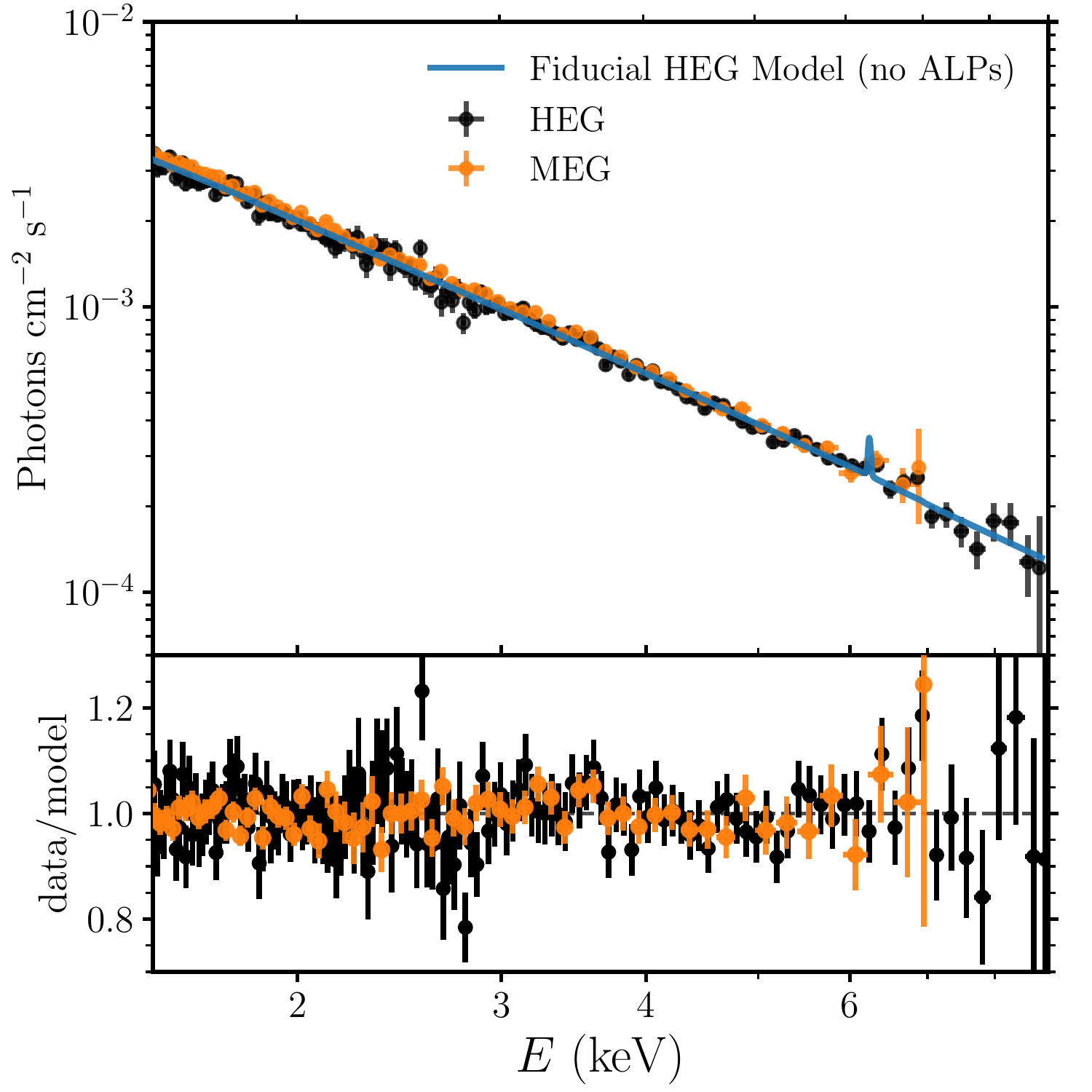}
    \caption{Top panel: The best-fitting fiducial model of the form \texttt{tbabs*tbpcf(pow+zgauss)} to the HEG data (blue) without ALPs, together with the combined HEG (black) and MEG (orange) data from the HETG observations of NGC~1275. The data are shown over the considered HEG energy range ($1.5-8.9$\,keV), although we include MEG data down to $1$\,keV in our analysis. Bottom: Fit residuals from the best fit model for each data set. The data have been rebinned for plotting purposes but all fits are performed on the unbinned data. The MEG best fit model is identical to the HEG bar minor differences in normalization (see e.g. Table~\ref{table:fiducial}).
    }
    \label{fig:spectra}
\end{figure}

We use the same Chandra High-Energy Transmission Grating (HETG) data as \citetalias{reynolds_astrophysical_2020} and \citet[][hereafter \citetalias{reynolds_probing_2021}]{reynolds_probing_2021}. Data from the HETG are split into two sets, corresponding to the high-energy grating (HEG) and medium-energy grating (MEG); the spectra are shown in Fig.~\ref{fig:spectra}. The observations form part of a Cycle-19 Large Project and were taken in 15 separate visits between 2017 October 24 and 2017 December 5 forming a total exposure of $490$\,ks (see \citetalias{reynolds_probing_2021} for further details of exposure times and precise dates). The actual reduced data used in this work are from \citetalias{reynolds_probing_2021}, who were able to slightly improve the background subtraction. \citetalias{reynolds_probing_2021} also describe an improved astrophysical model; they find that using a partially covering absorber not only improves the fit to the data, but also brings the inferred column density closer to the expected value from ALMA observations of the Perseus core \citep{nagai_alma_2019}. \cite{nagai_alma_2019} estimate an H$_2$ column density of $N_{H_2} \approx 2 \times 10^{22}\,{\rm cm}^{-2}$ from HCN and HCO$+$ absorption of the emission from the parsec-scale jet; this line-of-sight H$_2$ absorbing column is very difficult to reconcile with the observed X-ray power-law, which has a formal 90 per cent confidence limit of $N_H < 3 \times 10^{19}\,{\rm cm}^{-2}$ when fitted with a fully covering cold absorption model \citepalias{reynolds_probing_2021}. One physical interpretation is that the X-rays come from a composite source: an unabsorbed, compact corona associated with an accretion disc, and a heavily absorbed component associated with a parsec-scale jet, the latter of which contributes $\approx 15-20$ per cent of the X-ray continuum. The clumpy molecular gas in the cluster core is clumpy on similar scales to the jet \citep{nagai_alma_2019}, so this `partial covering' scenario is reasonable. Following \citetalias{reynolds_probing_2021}, we fit the \xspec\ model \texttt{tbabs*tbpcf(pow+zgauss)}, which includes partial covering absorption (\texttt{tbpcf}), Galactic absorption (\texttt{tbabs}) and a narrow Gaussian to model the $6.4$\,keV Fe line (\texttt{zgauss}). For the partial covering absorber, we find a best fit covering factor of $f_{\rm cov}=9.48 (\pm 2.85) \times10^{-2}$ and a column density $N_H=7.63 (\pm 2.24) \times10^{22}~{\rm cm}^{-2}$ when the Galactic column density in the \texttt{tbabs} model is kept fixed at $1.32 \times10^{21}$ \citep{kalberla_leidenargentinebonn_2005}. If the Galactic column density is allowed to vary a slightly higher covering factor for the \texttt{tbpcf} model of $f_{\rm cov}=1.75(\pm0.49) \times10^{-1}$ is favoured, with a Galactic column density of $1.71 (\pm 0.22) \times10^{21}$. We keep the Galactic density column fixed in this case, and adopt the best fit set of parameters given in Table~\ref{table:fiducial} as our baseline astrophysical model for the re-analysis. The best-fit model for the HEG data set is shown in Fig.~\ref{fig:spectra}, as are the fit residuals (without ALPs) for both the HEG and MEG data. 

\subsection{Statistical Procedure}
\label{sec:stats}
Our fitting procedure and statistical analysis follows the Bayesian procedure described by \citet[][see also \citetalias{reynolds_astrophysical_2020}; section 4 of \citealt{sisk_reynes_new_2021}]{marsh_new_2017}. We compute a grid of survival probability curves in $(m_a,\g)$ space with a range of random number seeds $i$ such that each ALP survival probability curve is defined by three variables $(m_a,\g,i)$. We compute curves in the range $\log_{10}(m_a / {\rm eV}) \in [-13.7,-10.5]$ and $\log_{10}(\g/{\rm GeV}^{-1}) \in [-13.0,-10.2]$ at $0.1$\,dex intervals. We follow \citetalias{reynolds_astrophysical_2020} in assuming that, at lower values of $\g$ and $m_a$, the ALP curves are statistically indistinguishable to those at the lower limits of our adopted calculation range -- an assumption we have checked -- so that results from $\log_{10}(m_a/{\rm eV})=-13.7$ can be extrapolated down to arbitrarily low $m_a$. We consider $N = 200$ field realisations in each case, which results in a library of $185,600$ survival curves for each of the five magnetic field models we consider. This value of $N$ is lower than the $500$ used by \citetalias{reynolds_astrophysical_2020}, but is necessary to prevent prohibitive computational cost given the increased number of field models, and that we must generate GRF realisations and consider smaller $\dz$ in some cases. Our choice of $N$ introduces some noise into the limits, but as we shall see similar results are recovered for comparable assumptions and it is still possible to distinguish systematic differences. 

We fit each model to the HEG and MEG spectra by combining the ALPs model with our baseline astrophysical model, such that we fit  \texttt{tbabs*ALPs(tbpcf*(pow+zgauss))}. For each ALP model, we minimise the \cite{cash_parameter_1979} statistic, or $C$-statistic, and record the best-fit (lowest) value. We then have a value of $C(m_a,\g,i)$  for each model realization, and can construct posterior probabilities of the form
\begin{equation}
{\cal P} (m_a,\g,i) \propto \exp (-C/2), 
\end{equation}
where we assume flat priors in $\ln m_a$ and $\ln \g$ over the range $\log_{10}(m_a / {\rm eV}) \in [-30.0,-11.1]$ and $\log_{10}(\g/{\rm GeV}^{-1}) \in [-19.0,-10.7]$. The posterior probabilities are normalized so that
\begin{equation}
    \sum_{m_a,\g,i} {\cal P} (m_a,\g,i) = 1.
\end{equation}
We then marginalize over the magnetic field realizations $i$, 
\begin{equation}
  {\cal P} (m_a,\g) = \sum_{i}^N {\cal P} (m_a,\g,i),
\end{equation}
to obtain a posterior probability at every point in $(m_a,\g)$ space, again assuming a flat prior on the field realisations. This marginalization step accounts for the `look-elsewhere effect' due to the unknown magnetic field structure along the line of sight. Limits at a given confidence level can than be calculated by sorting the points in ${\cal P} (m_a,\g)$ and finding the pairs of $(m_a,\g)$ with the highest ${\cal P} (m_a,\g)$ that cumulatively account for the required percentage of the total posterior probability (e.g., $\approx99.7$ per cent for a $3\sigma$ limit). We follow \cite{sisk_reynes_new_2021} in grouping pairs with identical ${\cal P}$ and assigning each pair the mean cumulative ${\cal P}$ of the group. This choice has only a very small cosmetic effect on the shape of the inferred limits. 

\begin{figure}
         \centering
         \includegraphics[width=\linewidth]{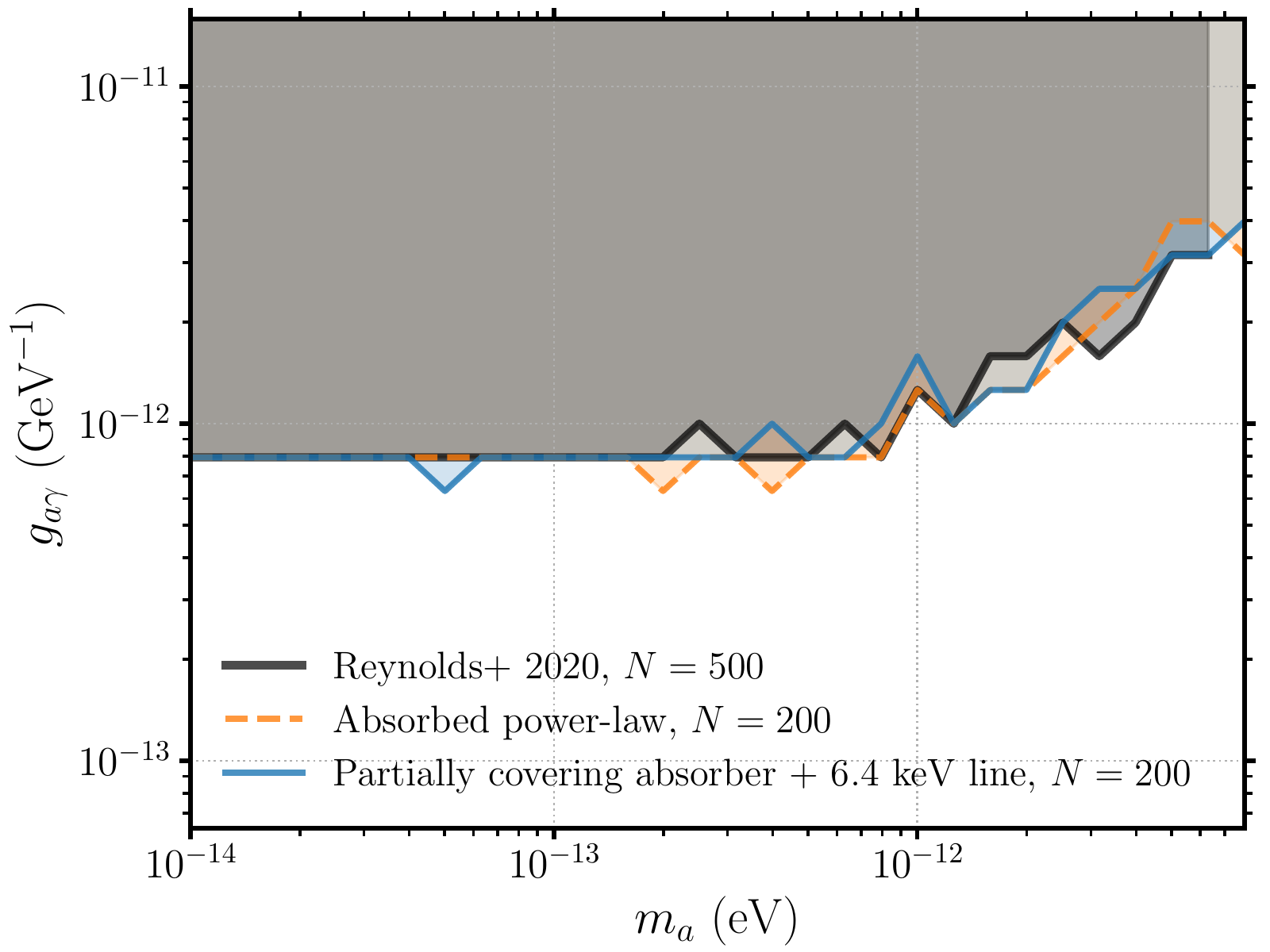}
     \caption{
     $99.7$ per cent limits obtained in this work using two different approaches for modelling the X-ray spectrum, zoomed in to the $10^{-14}\geq m_a/{\rm eV}< 10^{-11}$ region. The partially covering absorber fits use spectral models of the \xspec\ form \texttt{tbabs*tbpcf(ALPs*(pow+zgauss))}, whereas the absorbed power-law models use \texttt{phabs*zphabs(ALPs*pow)}. For comparison, we also show the limits from \citetalias{reynolds_astrophysical_2020} who used the absorbed power-law method. The limits are not very sensitive to the choice of spectral model, and we reproduce very similar results as  \citetalias{reynolds_astrophysical_2020} when using the equivalent spectral model. 
     }
    \label{fig:spectral_model_lims}
\end{figure}

\begin{figure*}
    \centering
    % \begin{subf
    \includegraphics[width=0.8\linewidth]{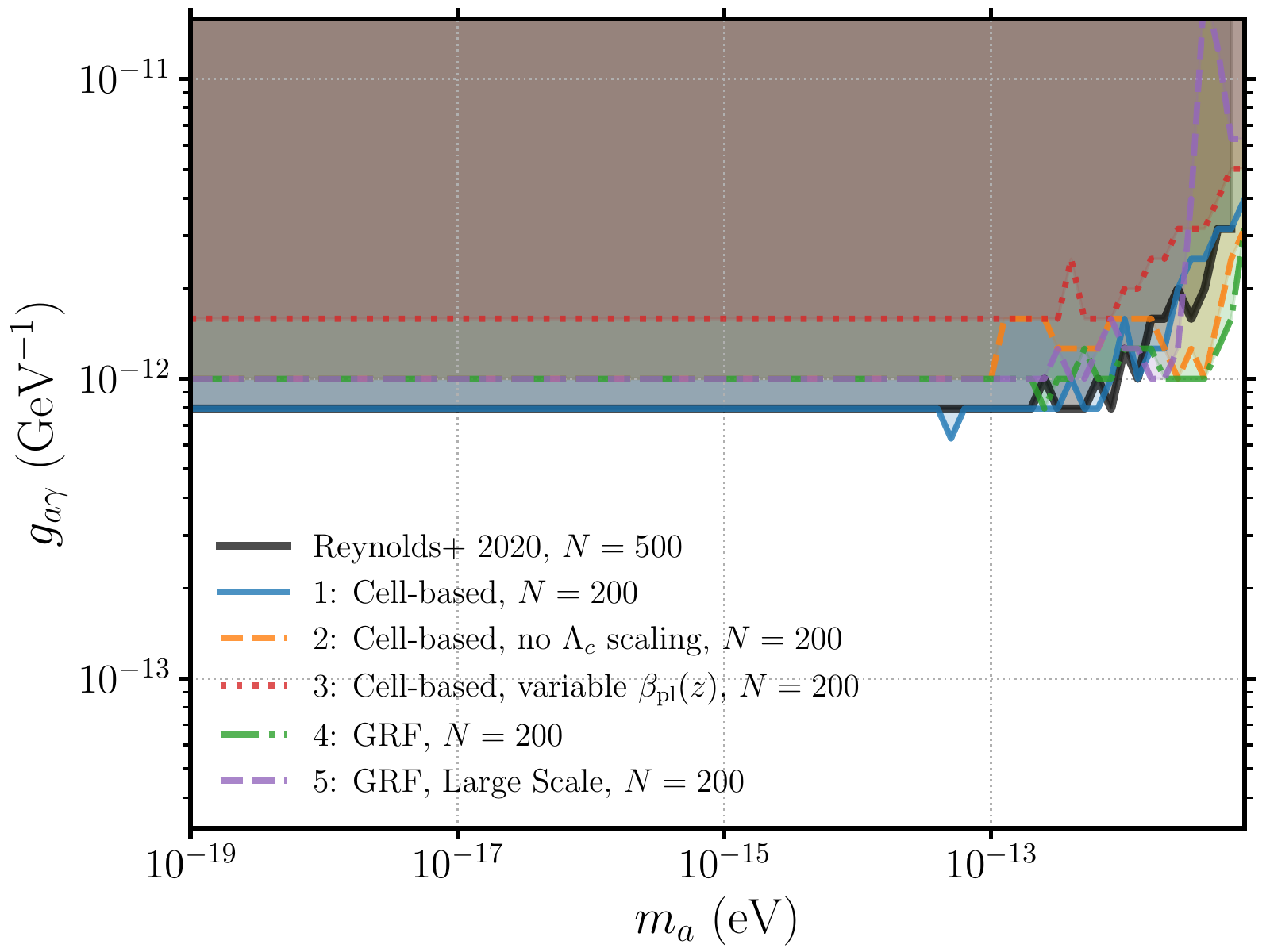}
    \caption{How does the choice of magnetic field model affect the NGC 1275 X-ray limits on light axion-like particles? A comparison of $99.7$ per cent ALP limits obtained using different approaches for modelling the magnetic field. The models used are given in Table~\ref{table:models} and the colours of the lines match those in the table and in Fig.~\ref{fig:Bfield_and_pgg}. The limits typically vary by 0.1 dex at $m_a<10^{-14}$~eV, with the most pessimistic model (model 3) resulting in weaker constraints on $\g$ by 0.3 dex. Overall, the limits obtained are not very sensitive to whether a cell-based or GRF approach is used, and the choices made about the coherence length of the magnetic field only change the limits by $0.1$\,dex.}
    \label{fig:lims_comparison}
\end{figure*}

\subsection{Limits on ALP parameters}
\label{sec:limits}
We calculate 99.7 per cent limits on ALP parameters using the above procedure and baseline spectral model, but now including ALPs.  We calculated limits with two different spectral models: the partially covering absorber models use models of the \xspec\ form \texttt{tbabs(ALPs*tbpcf(pow+zgauss))}, whereas the absorbed power-law models use \texttt{phabs*(ALPS*zphabs(pow))}. We also model the magnetic field in five different ways as described in section~\ref{sec:field_models} and as summarised in Table~\ref{table:models}. 

We begin by examining the limits obtained with the two different X-ray spectral models; these limits are shown in Fig.~\ref{fig:spectral_model_lims}, zoomed in to the $m_a\geq10^{-14}$~eV region and are also compared to the Model B limits from \citetalias{reynolds_astrophysical_2020}. The first result apparent from Fig.~\ref{fig:spectral_model_lims} is that we obtain very similar limits to \citetalias{reynolds_astrophysical_2020} when using our code \alpro, acting as an independent test of the \citetalias{reynolds_astrophysical_2020} results and showing that our new code behaves as expected. Without ALPs, the partially covering absorber plus $6.4$\,keV emission line model gives $C/{\rm dof} = 4857/4863$, compared to $C/{\rm dof}=4923/4865$ for the simple absorbed power-law model. This improvement in goodness-of-fit might be expected to give slightly tighter limits on $\g$. However, the limits obtained with the improved spectral model are extremely similar to those obtained with the simpler absorbed power-law, with near-identical results at low mass ($m_a \lesssim 10^{-13}$\,eV) and only small differences for $m_a \gtrsim 10^{-13}$\,eV, where the limits are in any case slightly noisy. We thus conclude that the limits are insensitive to the details of the approach used to model the X-ray spectrum of NGC 1275, as long as a physically sensible astrophysical model is used that adequately reproduces the data.

In Fig.~\ref{fig:lims_comparison}, we show the $99.7$ per cent limits obtained with the five different models for the magnetic field, now over a wider range in $m_a$. Here, we do see some diversity in the limits obtained. Models 2, 4 and 5 produce slightly weaker limits than \citetalias{reynolds_astrophysical_2020} at low mass ($m_a \lesssim 10^{-13}$\,eV), by $0.1$\,dex, ruling out $\g > 10^{-12}$\invgev\ at $99.7$ per cent confidence. Weaker limits from model 2, which does not scale the coherence length (or more accurately, the cell size $\Delta z$) with radius, would already be expected based on the mean survival probability curves shown in Fig.~\ref{fig:Bfield_and_pgg}. There, the mean survival probability from model 2 is significantly higher than in model 1, due only to the different scalings of $\Delta z$, and this translates directly into slightly weaker limits on $\g$. Since model 4 uses a GRF without any scaling of $\lc$ with $z$, the agreement with the equivalent cell-based model (model 2) shows that the exact choice of how to model the spatial structure of the field (cell-based versus GRF) is a sub-leading effect.   

The results from model 5, the `large-scale' GRF model, are particularly interesting. This model is designed to approximate larger scale magnetic field structures in the cluster that are coherent on large scales of $\gtrsim 50$\,kpc. \cite{libanov_impact_2020} have recently suggested that large-scale, coherent or `regular' field structures in the ICM might significantly weaken ALP limits using a similar method to ours, but applied to gamma-ray observations of NGC 1275. They use a uniform bubble model originally described by \cite{gourgouliatos_structure_2010}, with a maximum radius of $93$\,kpc. In our case, the limits do weaken slightly using the large-scale GRF model, but the effect is small. Our overall conclusion is that even if coherent large-scale ($\gtrsim 50$\,kpc) magnetic fields are present in Perseus, these do not necessarily significantly weaken the limits on $\g$ for low-mass ALPs. Some of these conclusions may be sensitive to the way we decided to model the magnetic field, and the result may be different with alternative data-sets; however, we stress that our stochastic model allows us to calculate the limits while still including the `look-elsewhere' effect and marginalising over random number seed, which was not accounted for in the \cite{libanov_impact_2020} analysis. Although we use a different dataset and waveband, our results suggest that the significantly weaker limits found by \cite{libanov_impact_2020} are specific to the field model adopted, rather than being a general feature of large-scale ICM magnetic fields that are coherent over $50-200$\,{\rm kpc}. 

Model 3, with the variable $\betapl(z)$, produces the least stringent limits on $\g$, weaker by $0.3$\,dex at low mass, and is a clear outlier. Inspecting the right-hand panel of Fig.~\ref{fig:Bfield_and_pgg}, this might initially seem surprising, since the conversion probability is comparable in amplitude to models 2 and 4. However, $\betapl$ increases to $\approx 630$ at $1$\,Mpc, translating to significantly decreased magnetic fields at large distances and a correspondingly smaller product $(B_\perp L)^2$. This leads to a small ALP signal at low $E$ in particular. Since the highest signal-to-noise ratio is obtained at low energies (closer to $1$\,keV) in both the HEG and MEG data, the limits are particularly sensitive to conversion probability this region, explaining the relatively weak limits for this variable $\betapl(z)$ model. 

At $m_a \gtrsim 10^{-12}$\,eV, the limits on $\g$ span a range of $\approx0.5$\,dex. Models 2 and 4, neither of which scale $\lc$ with radius, actually produce slightly more stringent limits at $m_a \gtrsim 10^{-12}$\,eV than obtained by \citetalias{reynolds_astrophysical_2020}. Conversely, the large-scale GRF model results in weaker limits in the same mass range. However, this section of the constraints plots is rather noisy in most models, perhaps due us needing to use $N=200$ realisations of the magnetic field (rather than, say, $N=500$ as used by \citetalias{reynolds_astrophysical_2020}). Generally speaking, the shape of the high--$m_a$ envelope of the constraints appears to be quite sensitive to the magnetic field model used, particularly the scale-lengths in the magnetic field model. Our results suggest that this region of parameter space -- where $m_a$ is comparable to the range of $\omegap$ in the cluster -- should be treated with caution when interpreting ALP limits from cluster-hosted AGN.

\begin{figure*}
    \centering
    % \begin{subf
    \includegraphics[width=1.0\linewidth]{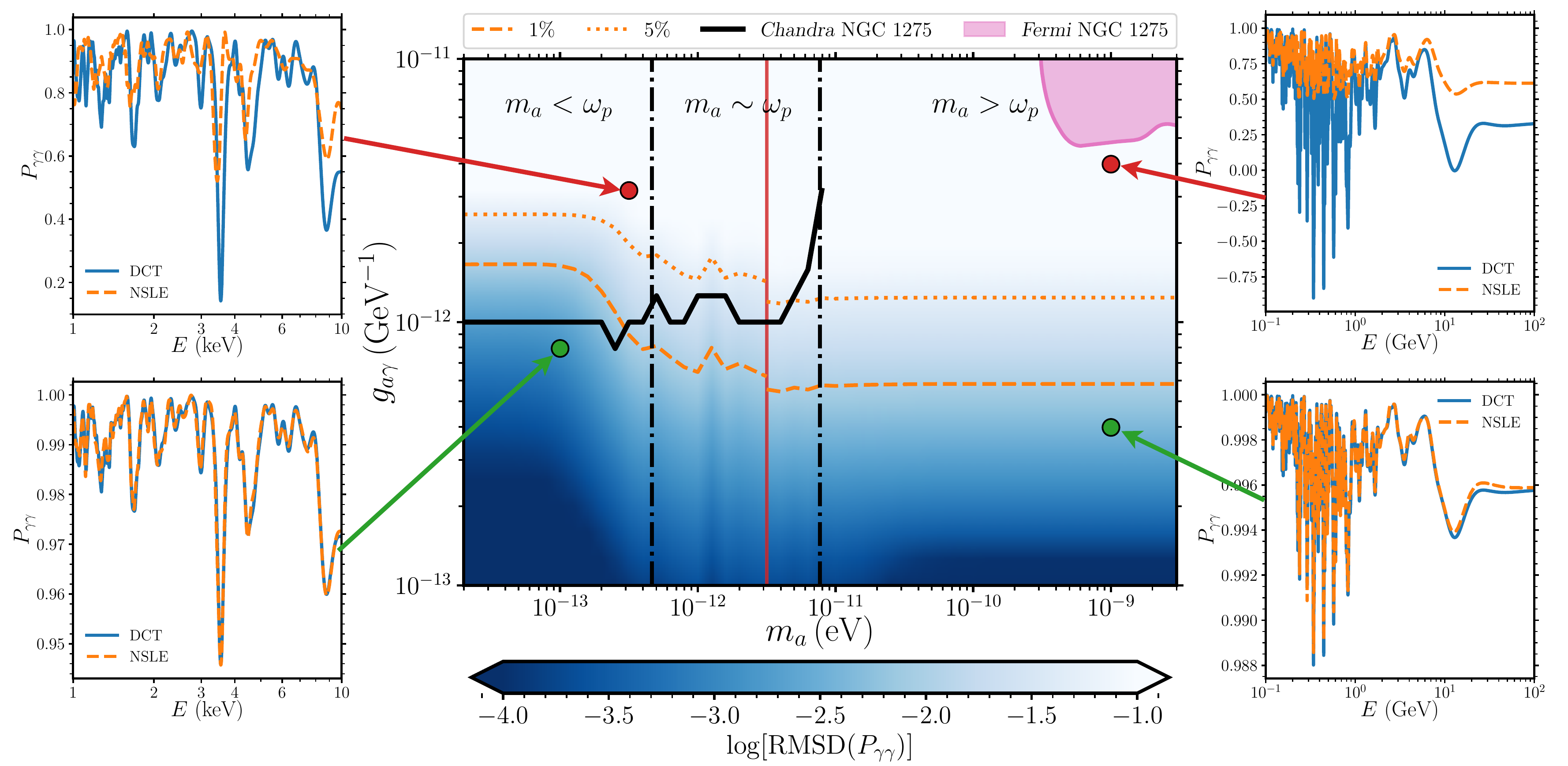}
    \caption{A demonstration of the application of the Fourier formalism to the Perseus cluster across a range of $(m_a,\g)$ parameter space, described in section~\ref{sec:fourier_app}. The colour-map in the central panel shows the the level of agreement between the DCT/Fourier and NSLE approaches, as measured by the logarithm of the root-mean-square deviation (RMSD), from equation~\ref{eq:rmsd}. The dashed (dotted) orange line marks the ${\rm RMSD}=0.01$ (${\rm RMSD}=0.05$)  contour, the vertical dot-dashed lines mark the range of $\omegap$ in our model for the Perseus cluster, and the vertical red line marks the value of $m_a$ below (above) which the massless (massive) formalism is used to calculate $\pgg$.  The NGC 1275 ALP limits from {\sl Fermi} \citep{ajello_search_2016} and {\sl Chandra} (our model 4) are also plotted. We show a comparison of the computed $\pgg$ for four different locations in parameter space spanning both the massless and massive ALP regime. Overall, the figure demonstrates the potential of the Fourier formalism and shows it can already be applied to data with comparable constraining power to those used to obtain X-ray limits from NGC 1275. 
    }
    \label{fig:fourier_demo}
\end{figure*}

\section{Discussion and Applications}
\label{sec:discuss}
\subsection{Application of the Fourier formalism}
\label{sec:fourier_app}
The Fourier formalism developed by \cite{marsh_fourier_2022} and described briefly in section~\ref{sec:fourier} can be applied to the problem of photon-ALP conversion in the Perseus cluster. \cite{marsh_fourier_2022} already presented results from a model appropriate for Perseus, using a simple single $\beta$--law (equation~\ref{eq:betalaw}) for the massless ALP case and neglecting $n_e$ in the massive ALP case. Here, we repeat similar calculations using discrete cosine transforms (DCTs) across a wide range of parameter space and record the level of agreement with the more general numerical Schr{\" o}dinger-like equation (NSLE) solution. We consider both the massive and massless regimes, which are determined by the range of $\omegap$ in the Perseus cluster. For the purposes of this application we neglect the `general' case described by \cite{marsh_fourier_2022}, in which a resonant point (at which $m_a = \omegap$) is crossed along the path $z$.

In the massless case, where $m_a < \omegap$, it is necessary to calculate a phase $\varphi$, defined as 
\begin{equation}
    \varphi(z) = \frac{1}{2} \int_0^z \omegap(z^\prime) dz^\prime.
\end{equation}
As with a single $\beta$-law, $\varphi$ can be calculated analytically for equation~\ref{eq:ne}, but we use a numerical Simpson integration for generality. For a single GRF field realisation ($i=0$), and for both $x$ and $y$ polarization, we compute the function $G(\varphi)= \g B_x / \omegap^2$ and then calculate $c_G$, the autocorrelation function of $G$ in $\varphi$-space. We then
calculate the conversion probability $\pgax$ by taking the DCT of $c_G$ with $N_{\rm dct}=10^5$ Fourier samples and a conjugate variable $1/E$. In the massive case, we essentially repeat the above exercise but without the calculation of $\varphi$, instead calculating the autocorrelation function of the magnetic field in real space, $c_{B_x}$, and taking the DCT of this to obtain $\pgax$, with conjugate variable $\eta=m_a^2/(2E)$. Once $\pgax$ and $\pgay$ are known, the unpolarized survival probability $\pgg$ follows from equation~\ref{eq:unpol}. The full procedure in both cases is described in detail by \cite{marsh_fourier_2022}.

We focus on X-ray energies ($1-10$\,keV) in the massless ALP case. In the massive ALP regime the energy range depends on $m_a$ (from the definition of $\eta$). We consider an energy range $0.1-100\,(m_a/10^{-9})^2$\,GeV, which is broadly appropriate for the NGC 1275 {\sl Fermi}-derived limits from \cite{ajello_search_2016}. We evaluate the agreement with the NSLE method by recording the normalised root-mean-square deviation, defined as 
\begin{equation}
    {\rm RMSD} (\pgg) = \frac{1}{\bar{P}_{\gamma\gamma}^{\rm NSLE}}
    \sqrt{\frac{\sum^{N_E}_j \left(P_{\gamma\gamma,j}^{\rm NSLE} - P_{\gamma\gamma,j}^{\rm DCT} \right)^2}{N_E}},
\label{eq:rmsd}
\end{equation}
where the $j$ index denotes the energy bin, $\bar{P}_{\gamma\gamma}$ denotes a mean survival probability averaged over $N_E$ energy bins, and the NSLE and DCT superscripts refer to the method used to calculate $\pgg$.

In the central panel of Fig.~\ref{fig:fourier_demo}, the logarithm of  ${\rm RMSD} (\pgg)$  is plotted as a colourmap as a function of $m_a$ and $\g$ and the one and five per cent (${\rm RMSD} = 0.01,0.05$) contours are marked. We compute the survival probability across a wide range of $m_a$ and $\g$, using the massless formalism for $\log m_a < -11.5$ and the massive formalism for $\log m_a > -11.5$. We show four representative examples of survival probability curves in relevant energy ranges, chosen so that two show good agreement and two do not. The limits from model 4 in this work, and the NGC 1275 {\sl Fermi} limits from  \cite{ajello_search_2016} are also plotted, and the vertical dot-dashed lines mark the range of $\omegap$ (at distances $10-1800$\,kpc from the cluster centre) in natural (eV) units,  as calculated from equation~\ref{eq:ne}.  If we take the one per cent contour of ${\rm RMSD}$ as constituting reasonable agreement, then the Fourier formalism can already be applied to calculations in the regime of $\g$ probed by X-ray observations in the massless ALP regime. In fact, the formalism is likely to give identical statistical results for higher ${\rm RMSD}$ values comparable to the residuals in the data. The {\sl Fermi} limits lie at significantly higher $\g$, where the perturbative calculation breaks down, so the Fourier formalism cannot yet be applied to gamma-ray ALP searches. However, the scheme still offers potential for the future if the constraining power in the GeV regime can be improved by a factor of 10 or so. 

By making use of FFT techniques for the DCT, the Fourier formalism can convey a significant performance advantage, but, as discussed by \cite{marsh_fourier_2022}, this speed up in the calculation depends somewhat on the calculation considered. Using \alpro, calculating the $\pgg$ curve shown in the bottom right of Fig~\ref{fig:fourier_demo} takes $\approx3.9$\,s with the NLSE approach with $N_E=1000$ energy bins, compared to the Fourier/DCT approach which takes $\approx0.7$\,s for $N_{\rm dct}=10^5$. As $N_E \to {\cal N}_{\rm dct}$, the speed advantage of the Fourier/DCT approach  improves linearly with $N_E$, and can be dramatic. The Fourier formalism is therefore likely to be most useful for high energy resolution or wide energy bands, but even at more modest energy resolutions it already can give extremely fast results for reasonably complex models. The Fourier formalism may also be used to infer information about ALPs and magnetic field structure directly from the residuals in X-ray gamma-ray data. As noted by \citealt{marsh_fourier_2022}, ALP-induced irregularities encode the autocorrelation function of the line-of-sight magnetic field and it may be possible to map directly from the data residuals to this function \citep[see also][]{conlon_improving_2019,kachelriess_origin_2021}. The scheme thus offers potential for the future, particularly as a tool for ALP searches with the next-generation {\sl Athena} X-ray telescope \citep{nandra_hot_2013,conlon_projected_2018}.

\label{sec:performance}

\subsection{Future work, limitations and implications for other clusters}
\label{sec:other_limits}

We have chosen to focus on the Perseus cluster specifically, but our results clearly have implications for ALP limits derived from observations of X-ray bright AGN in and behind other clusters. The limits derived so far in the literature form an inhomogeneous data set with slightly different assumptions in each study. In each case the availability of RM data and accurate density and pressure profiles varies, while there may be expected to be some intrinsic inter-cluster variability in, for example, the magnitude of $\betapl$, or the coherence of the magnetic field. Put simply, we have shown that five magnetic field models that produce similar magnetic pressures in the inner regions of the cluster and a comparable distributions of RMs -- both of which broadly consistent with observations under reasonable assumptions -- can produce different strength ALP signals. One interpretation is that the scatter in the limits in Fig.~\ref{fig:lims_comparison} act to crudely encode the systematic uncertainty of the various assumptions in these field models. Based on this reasoning, we might expect that uncertainty on the strength and detailed radial profile of ICM magnetic fields introduces a systematic uncertainty of around $0.3$\,dex into the various astrophysical limits on ALPs. Moving forward, it would be useful to have a more uniform set of observational constraints on the clusters used for ALP studies to date, through a targeted combination of high-sensitivity RM studies and X-ray observations.

A fruitful avenue for future work would be to use magnetohydrodynamic (MHD) simulations of cluster environments \citep[e.g.][]{xu_turbulence_2009,donnert_cluster_2009,beresnyak_turbulent_2016,vazza_amplification_2014,vazza_resolved_2018} to inform studies such as ours. MHD simulations allow one to study the evolution of magnetic fields in a dynamic environment and the resulting impact on their coherence, strength and overall structure. In addition, magnetic fields may be anisotropic, whereas the GRF model we used assumed isotropy, and ALP signals could be predicted directly from the simulation outputs. Including further insights from MHD simulations and, more generally, gaining a better understanding of the physical processes that govern the magnetised ICM will be critical for the future of cluster-based ALP studies. One important quantity to constrain -- observationally and through MHD  modelling -- is the value of $\betapl$ in the outer regions of the cluster, since we have shown that the radial profile of $\betapl$ does have an impact on the  photon-ALP conversion and resulting limits obtained.

\section{Conclusions}
\label{sec:conclusions}
We have revisited the problem of photon-axion conversion in the Perseus cluster magnetic field with NGC 1275 as a background source. We have re-analysed the {\sl Chandra} X-ray data, adopting different approaches to modelling the magnetic field and an improved spectral model. The main result of our work is that for well-motivated turbulent field models the limits on $(m_a,\g)$ obtained by \cite{reynolds_astrophysical_2020} are, in general, quite insensitive to these choices. Overall, the result that $\g<8\times10^{-13}$\invgev\ for $m_a < 10^{-12}$\,eV\ is robust under the assumption that the ratio of thermal to magnetic pressure in the Perseus cluster is $\betapl\approx100$. Our other main conclusions are given below. 
\begin{itemize}
    \item We review the evidence for turbulent magnetic fields in Perseus. We conclude that turbulent fields are likely, and that Model B of \citetalias{reynolds_astrophysical_2020} is a reasonable approximation to the magnetic field in Perseus cluster under the justified assumption that $\betapl \approx 100$. Model A from \citetalias{reynolds_astrophysical_2020} predicts a pressure-profile in excess of that observed for $\betapl=100$ (equivalently, it predicts a $\betapl$ that is lower than expected in Perseus and other cool-core clusters).  
    \item We conduct a sensitivity study using a Gaussian random field with the same radial profile as Model B from \citetalias{reynolds_astrophysical_2020}. We examine the sensitivity of the ALP signature to the resolution of the photon-ALP simulations, $\dz$, and discuss with reference to the scale lengths associated with the turbulence. For a Kolmogorov field we find that convergence is generally reached at scales below the coherence length, around the minimum scale length of the turbulence, and that under-resolving the magnetic field can lead to over-estimates of the ALP signal for a given $\g$ and $\myvector{B}(z)$.
    \item Informed by the sensitivity study, we re-analyse the NGC 1275 Chandra X-ray data. We use an improved data reduction and X-ray spectral model that accounts for a composite X-ray source surrounded by a clumpy, partially covering absorber. We confirm the basic results from
    \cite{reynolds_astrophysical_2020} and find that the limits derived are insensitive to the X-ray spectral model, as long as the spectral models are appropriate and sufficient to describe the data.
    \item We re-derive limits using the same X-ray data on NGC 1275 for five different magnetic field models. At low $m_a$ ($m_a\lesssim10^{-13}$\,eV), marginally weaker limits on $\g$ (by $0.1$\,dex) are obtained with different magnetic field models, including a Gaussian random field model designed to approximate  kpc-scale turbulence. Our most pessimistic model, which has a ratio of thermal to magnetic pressure that increases to $\betapl \approx 850$ by the virial radius, predicts weaker limits by $0.3$\,dex at low $m_a$. We conclude that the limits are largely insensitive to whether a cell-based or GRF approach is used to model the magnetic field, or the choices made about the coherence length, but systematic uncertainties relating to the the magnitude and radial profile of the magnetic field strength (or, equivalently, $\betapl$) persist.
    \item Using a `large-scale' GRF model, with magnetic fields that are coherent on $\gtrsim 50\,{\rm kpc}$ scales, only causes the limits at low $m_a$ to weaken by $0.1$\,dex. Our work suggests that the significantly weaker limits found by \cite{libanov_impact_2020} are not a general feature of large-scale ICM magnetic fields, and are instead specific to the particular realisation of the magnetic field adopted. 
    \item We show that the new Fourier formalism described by \cite{marsh_fourier_2022} can be applied to large regions of the relevant parameter space for a study such as ours and can result in a significant performance improvement when computing photon-ALP survival probability curves. 
    \item We introduce our new Python package \alpro\ for solving the Schr{\" o}dinger-like equation for ALP-photon propagation. \alpro\ also an implementation of the Fourier formalism described by \cite{marsh_fourier_2022}. The code is publicly available at \url{https://github.com/jhmatthews/alpro} with accompanying documentation and is used for all numerical photon-ALP survival probability calculations in this work. 
\end{itemize}
Overall, our work suggests systematic uncertainties in the magnetic field strength and structure along the line of sight remain important to understand. To make progress, further high-sensitivity RM observations across larger regions of the cluster would be extremely valuable. It is also important to develop a better theoretical and observational understanding of the value of the plasma-beta ($\betapl$) in the outer regions of clusters in general. Despite these uncertainties, X-ray observations of cluster-hosted AGN remain one of the most promising and important tools for constraining the properties of very light ALPs. 

\section*{Acknowledgements}
We thank an anonymous referee for a constructive and helpful report. J.H.M acknowledges a Herchel Smith Fellowship at Cambridge. C.S.R. thanks the STFC for support under the Consolidated Grant ST/S000623/1, as well as the European Research Council (ERC) for support under the European Union’s Horizon 2020 research and innovation programme (grant 834203). J.S.R acknowledges support from the Science and Technology Facilities Council (STFC) under grant ST/V50659X/1 (project reference 2442592).  D.M. is supported by the European Research Council under Grant No. 742104 and by the Swedish Research Council (VR) under grants 2018-03641 and 2019-02337. P.E.R. thanks the Gates Cambridge Trust for supporting her doctoral studies. This work was supported, in whole or in part, by the Bill \& Melinda Gates Foundation (OPP1144). Under the grant conditions of the Foundation, a Creative Commons Attribution 4.0 Generic License has already been assigned to the Author Accepted Manuscript version that might arise from this submission. The scientific results reported in this article are based in part on observations made by the {\sl Chandra} X-ray Observatory and published previously in cited articles. This work was performed using resources provided by the Cambridge Service for Data Driven Discovery (CSD3) operated by the University of Cambridge Research Computing Service (www.csd3.cam.ac.uk), provided by Dell EMC and Intel using Tier-2 funding from the Engineering and Physical Sciences Research Council (capital grant EP/P020259/1), and DiRAC funding from the Science and Technology Facilities Council (www.dirac.ac.uk). We would like to thank Pierluca Carenza, Jamie Davies, Richard Booth and Andy Fabian for helpful discussions, and Will Alston for answering some questions about \xspec. 

\software{We gratefully acknowledge the use of the following software packages: astropy \citep{astropy2013,astropy2018}, matplotlib \citep{matplotlib}, numba \citep{numba}, scipy \citep{2020SciPy-NMeth}, pandas \citep{mckinney-proc-scipy-2010,reback2020pandas}, OpenMPI \citep{openmpi} and \xspec\ \citep{arnaud_xspec_1996}. This paper includes results from \alpro\ version 1.0 \citep{alpro}, but versions 1.1 and later should be used to access the Fourier scheme and more complete documentation.}

%%%%%%%%%%%%%%%%%%%% REFERENCES %%%%%%%%%%%%%%%%%%

\label{lastpage}

\begin{thebibliography}{}
\makeatletter
\relax
\def\apjm@urlcharsother{\let\do\@makeother \do\$\do\&\do\#\do\^\do\_\do\%\do\~}
\def\apjm@doi{\begingroup\apjm@urlcharsother \@ifnextchar [ {\apjm@doi@}
  {\apjm@doi@[]}}
\def\apjm@doi@[#1]#2{\def\@tempa{#1}\ifx\@tempa\@empty \href
  {http://dx.doi.org/#2} {doi:#2}\else \href {http://dx.doi.org/#2} {#1}\fi
  \endgroup}
\def\apjm@eprint#1#2{\apjm@eprint@#1:#2::\@nil}
\def\apjm@eprint@arXiv#1{\href {http://arxiv.org/abs/#1} {{\tt arXiv:#1}}}
\def\apjm@eprint@dblp#1{\href {http://dblp.uni-trier.de/rec/bibtex/#1.xml}
  {dblp:#1}}
\def\apjm@eprint@#1:#2:#3:#4\@nil{\def\@tempa {#1}\def\@tempb {#2}\def\@tempc
  {#3}\ifx \@tempc \@empty \let \@tempc \@tempb \let \@tempb \@tempa \fi \ifx
  \@tempb \@empty \def\@tempb {arXiv}\fi \@ifundefined
  {mn@eprint@\@tempb}{\@tempb:\@tempc}{\expandafter \expandafter \csname
  mn@eprint@\@tempb\endcsname \expandafter{\@tempc}}}

\bibitem[\protect\citeauthoryear{Ajello et~al.,}{Ajello
  et~al.}{2016}]{ajello_search_2016}
Ajello M.,  et~al., 2016, \apjm@doi [\prl]
  {10.1103/PhysRevLett.116.161101}, 116, 161101

\bibitem[\protect\citeauthoryear{Aleksić et~al.,}{Aleksić
  et~al.}{2012}]{aleksic_constraining_2012}
Aleksić J.,  et~al., 2012, \apjm@doi [\aap]
  {10.1051/0004-6361/201118502}, 541, A99

\bibitem[\protect\citeauthoryear{Allen, Ettori  \& Fabian}{Allen
  et~al.}{2001}]{allen_chandra_2001}
Allen S.~W.,  Ettori S.,   Fabian A.~C.,  2001, \apjm@doi [\mnras] {10.1046/j.1365-8711.2001.04318.x}, 324, 877

\bibitem[\protect\citeauthoryear{Angus, Conlon, Marsh, Powell  \&
  Witkowski}{Angus et~al.}{2014}]{angus_soft_2014}
Angus S.,  Conlon J.~P.,  Marsh M. C.~D.,  Powell A.~J.,   Witkowski L.~T.,
  2014, \apjm@doi [\jcap] {10.1088/1475-7516/2014/09/026}, 09, 026

\bibitem[\protect\citeauthoryear{Arias, Jaeckel, Redondo  \& Ringwald}{Arias
  et~al.}{2010}]{arias_optimizing_2010}
Arias P.,  Jaeckel J.,  Redondo J.,   Ringwald A.,  2010, \apjm@doi [\prd] {10.1103/PhysRevD.82.115018}, 82, 115018

\bibitem[\protect\citeauthoryear{Arias, Cadamuro, Goodsell, Jaeckel, Redondo
  \& Ringwald}{Arias et~al.}{2012}]{arias_wispy_2012}
Arias P.,  Cadamuro D.,  Goodsell M.,  Jaeckel J.,  Redondo J.,   Ringwald A.,
  2012, \apjm@doi [\jcap]
  {10.1088/1475-7516/2012/06/013}, 2012, 013

\bibitem[\protect\citeauthoryear{Arik et~al.,}{Arik
  et~al.}{2009}]{arik_probing_2009}
Arik E.,  et~al., 2009, \apjm@doi [\jcap] {10.1088/1475-7516/2009/02/008}, 2009, 008

\bibitem[\protect\citeauthoryear{Armengaud et~al.,}{Armengaud
  et~al.}{2014}]{armengaud_conceptual_2014}
Armengaud E.,  et~al., 2014, \apjm@doi [JInst]
  {10.1088/1748-0221/9/05/T05002}, 9, T05002

\bibitem[\protect\citeauthoryear{Arnaud}{Arnaud}{1996}]{arnaud_xspec_1996}
Arnaud K.~A., 1996, ASPC, 101, 17

\bibitem[\protect\citeauthoryear{{Astropy Collaboration} et~al.,}{{Astropy
  Collaboration} et~al.}{2013}]{astropy2013}
{Astropy Collaboration} et~al., 2013, \apjm@doi [\aap]
  {10.1051/0004-6361/201322068}, \href
  {https://ui.adsabs.harvard.edu/abs/2013A&A...558A..33A} {558, A33}

\bibitem[\protect\citeauthoryear{{Astropy Collaboration} et~al.,}{{Astropy
  Collaboration} et~al.}{2018}]{astropy2018}
{Astropy Collaboration} et~al., 2018, \apjm@doi [\aj] {10.3847/1538-3881/aabc4f},
  \href {https://ui.adsabs.harvard.edu/abs/2018AJ....156..123A} {156, 123}

\bibitem[\protect\citeauthoryear{Balbus}{2000}]{balbus2000} Balbus S.~A., 2000, \apjm@doi[\apj]{10.1086/308732}, 534, 420

\bibitem[\protect\citeauthoryear{Balbus}{2001}]{balbus2001} Balbus S.~A., 2001, \apjm@doi[\apj]{10.1086/323875}, 562, 909 

\bibitem[\protect\citeauthoryear{Balbus \& Reynolds}{2010}]{balbus2010} Balbus S.~A., Reynolds C.~S., 2010, \apjm@doi[\apjl]{10.1088/2041-8205/720/1/L97}, 720, L97

\bibitem[\protect\citeauthoryear{Ballou et~al.,}{Ballou
  et~al.}{2014}]{ballou_latest_2014}
Ballou R.,  et~al., 2014, arXiv e-prints, p. arXiv:1410.2566

\bibitem[\protect\citeauthoryear{Beheshtipour, Krawczynski, \& Malzac}{Beheshtipour et~al.}{2017}]{beheshtipour2017} Beheshtipour B., Krawczynski H., Malzac J., 2017,  \apjm@doi [\apj] {10.3847/1538-4357/aa906a}, 850, 14. 

\bibitem[\protect\citeauthoryear{Beresnyak \& Miniati}{Beresnyak \&
  Miniati}{2016}]{beresnyak_turbulent_2016}
Beresnyak A.,  Miniati F.,  2016, \apjm@doi [\apj]
  {10.3847/0004-637X/817/2/127}, 817, 127

\bibitem[\protect\citeauthoryear{Berg, Conlon, Day, Jennings, Krippendorf,
  Powell  \& Rummel}{Berg et~al.}{2017}]{berg_constraints_2017}
Berg M.,  Conlon J.~P.,  Day F.,  Jennings N.,  Krippendorf S.,  Powell A.~J.,
   Rummel M.,  2017, \apjm@doi [\apj]
  {10.3847/1538-4357/aa8b16}, 847, 101

\bibitem[\protect\citeauthoryear{Bonafede, Feretti, Murgia, Govoni, Giovannini,
  Dallacasa, Dolag  \& Taylor}{Bonafede et~al.}{2010}]{bonafede_coma_2010}
Bonafede A.,  Feretti L.,  Murgia M.,  Govoni F.,  Giovannini G.,  Dallacasa
  D.,  Dolag K.,   Taylor G.~B.,  2010, \apjm@doi [\aap]
  {10.1051/0004-6361/200913696}, 513, A30

\bibitem[\protect\citeauthoryear{Bonafede et~al.,}{Bonafede
  et~al.}{2014}]{bonafede_giant_2014}
Bonafede A.,  et~al., 2014, \apjm@doi [\mnras] {10.1093/mnrasl/slu110}, 444, L44

\bibitem[\protect\citeauthoryear{Brockway, Carlson  \& Raffelt}{Brockway
  et~al.}{1996}]{brockway_sn_1996}
Brockway J.~W.,  Carlson E.~D.,   Raffelt G.~G.,  1996, \apjm@doi [Phys. Lett. B] {10.1016/0370-2693(96)00778-2}, 383, 439

\bibitem[\protect\citeauthoryear{Böhringer, Chon  \& Kronberg}{Böhringer
  et~al.}{2016}]{bohringer_cosmic_2016}
Böhringer H.,  Chon G.,   Kronberg P.~P.,  2016, \apjm@doi [\aap] {10.1051/0004-6361/201628873}, 596, A22

\bibitem[\protect\citeauthoryear{{CAST Collaboration}, Andriamonje  \&
  {others}}{{CAST Collaboration}
  et~al.}{2007}]{cast_collaboration_improved_2007}
{CAST Collaboration} Andriamonje S.,   {others} 2007, \apjm@doi [\jcap]
  {10.1088/1475-7516/2007/04/010}, 0704, 010

\bibitem[\protect\citeauthoryear{Carilli \& Taylor}{Carilli \&
  Taylor}{2002}]{carilli_cluster_2002}
Carilli C.~L.,  Taylor G.~B.,  2002, \apjm@doi [\araa] {10.1146/annurev.astro.40.060401.093852}, 40, 319

\bibitem[\protect\citeauthoryear{Cash}{Cash}{1979}]{cash_parameter_1979}
Cash W.,  1979, \apjm@doi [\apj] {10.1086/156922}, 228, 939

\bibitem[\protect\citeauthoryear{Chadha-Day, Ellis  \& Marsh}{Chadha-Day
  et~al.}{2021}]{chadha-day_axion_2021}
Chadha-Day F.,  Ellis J.,   Marsh D. J.~E.,  2021, arXiv e-prints, 2105,
  arXiv:2105.01406

\bibitem[\protect\citeauthoryear{Cheng}{Cheng}{1988}]{cheng_strong_1988}
Cheng H.~Y.,  1988, \apjm@doi [Physics Reports] {10.1016/0370-1573(88)90135-4},
  158, 1

\bibitem[\protect\citeauthoryear{Churazov, Forman, Jones  \&
  Böhringer}{Churazov et~al.}{2003}]{churazov_xmm-newton_2003}
Churazov E.,  Forman W.,  Jones C.,   Böhringer H.,  2003, \apjm@doi [\apj] {10.1086/374923}, 590, 225

\bibitem[\protect\citeauthoryear{Clarke, Kronberg  \& Böhringer}{Clarke
  et~al.}{2001}]{clarke_new_2001}
Clarke T.~E.,  Kronberg P.~P.,   Böhringer H.,  2001, \apjm@doi [\apj] {10.1086/318896}, 547, L111

\bibitem[\protect\citeauthoryear{Collaboration et~al.,}{Collaboration
  et~al.}{2013}]{planck_collaboration_planck_2013}
Collaboration P.,  et~al., 2013, \apjm@doi [\aap]
  {10.1051/0004-6361/201220040}, 550, A131

\bibitem[\protect\citeauthoryear{Conlon \& Rummel}{Conlon \&
  Rummel}{2019}]{conlon_improving_2019}
Conlon J.~P.,  Rummel M.,  2019, \apjm@doi [\mnras]
  {10.1093/mnras/stz211}, 484, 3573

\bibitem[\protect\citeauthoryear{Conlon, Day, Jennings, Krippendorf  \&
  Rummel}{Conlon et~al.}{2017}]{conlon_constraints_2017}
Conlon J.~P.,  Day F.,  Jennings N.,  Krippendorf S.,   Rummel M.,  2017,
  \apjm@doi [\jcap]
  {10.1088/1475-7516/2017/07/005}, 2017, 005

\bibitem[\protect\citeauthoryear{Conlon, Day, Jennings, Krippendorf  \&
  Muia}{Conlon et~al.}{2018}]{conlon_projected_2018}
Conlon J.~P.,  Day F.,  Jennings N.,  Krippendorf S.,   Muia F.,  2018, \apjm@doi
  [\mnras] {10.1093/mnras/stx2652},
  473, 4932

\bibitem[\protect\citeauthoryear{Davies, Meyer  \& Cotter}{Davies
  et~al.}{2020}]{davies_relevance_2020}
Davies J.,  Meyer M.,   Cotter G.,  2020, arXiv e-prints, 2011,
  arXiv:2011.08123

\bibitem[\protect\citeauthoryear{Day \& Krippendorf}{2018}]{day2018} Day F., Krippendorf S., 2018, \apjm@doi[Galax]{doi:10.3390/galaxies6020045}, 6, 45. 

\bibitem[\protect\citeauthoryear{Day \& Krippendorf}{2020}]{day_accelerating2020} 
Day F., Krippendorf S., 2020, \apjm@doi[\jcap]{doi:10.1088/1475-7516/2020/03/046}, 2020, 046. 

\bibitem[\protect\citeauthoryear{Donnert, Dolag, Lesch  \& Müller}{Donnert
  et~al.}{2009}]{donnert_cluster_2009}
Donnert J.,  Dolag K.,  Lesch H.,   Müller E.,  2009, \apjm@doi [\mnras] {10.1111/j.1365-2966.2008.14132.x}, 392,
  1008

\bibitem[\protect\citeauthoryear{Donnert, Vazza, Brüggen  \& ZuHone}{Donnert
  et~al.}{2018}]{donnert_magnetic_2018}
Donnert J.,  Vazza F.,  Brüggen M.,   ZuHone J.,  2018, \apjm@doi [\ssr] {10.1007/s11214-018-0556-8}, 214, 122

\bibitem[\protect\citeauthoryear{Ehret et~al.,}{Ehret
  et~al.}{2009}]{ehret_resonant_2009}
Ehret K.,  et~al., 2009, \apjm@doi [NIMPA] {10.1016/j.nima.2009.10.102}, 612, 83

\bibitem[\protect\citeauthoryear{Ensslin \& Vogt}{Ensslin \&
  Vogt}{2003}]{ensslin_magnetic_2003}
Ensslin T.~A.,  Vogt C.,  2003, \apjm@doi [\aap]
  {10.1051/0004-6361:20030172}, 401, 835

\bibitem[\protect\citeauthoryear{Enßlin \& Vogt}{Enßlin \&
  Vogt}{2006}]{enslin_magnetic_2006}
Enßlin T.~A.,  Vogt C.,  2006, \apjm@doi [\aap]
  {10.1051/0004-6361:20053518}, 453, 447

\bibitem[\protect\citeauthoryear{Fabian et~al.,}{Fabian
  et~al.}{2000}]{fabian_chandra_2000}
Fabian A.~C.,  et~al., 2000, \apjm@doi [\mnras] {10.1046/j.1365-8711.2000.03904.x}, 318, L65

\bibitem[\protect\citeauthoryear{Fabian, Sanders, Taylor, Allen, Crawford,
  Johnstone  \& Iwasawa}{Fabian et~al.}{2006}]{fabian_very_2006}
Fabian A.~C.,  Sanders J.~S.,  Taylor G.~B.,  Allen S.~W.,  Crawford C.~S.,
  Johnstone R.~M.,   Iwasawa K.,  2006, \apjm@doi [\mnras] {10.1111/j.1365-2966.2005.09896.x}, 366, 417

\bibitem[\protect\citeauthoryear{Feretti, Dallacasa, Giovannini  \&
  Tagliani}{Feretti et~al.}{1995}]{feretti_magnetic_1995}
Feretti L.,  Dallacasa D.,  Giovannini G.,   Tagliani A.,  1995, \aap, 302, 680

\bibitem[\protect\citeauthoryear{Feretti, Dallacasa, Govoni, Giovannini, Taylor
   \& Klein}{Feretti et~al.}{1999}]{feretti_radio_1999}
Feretti L.,  Dallacasa D.,  Govoni F.,  Giovannini G.,  Taylor G.~B.,   Klein
  U.,  1999, \aap, 344, 472

\bibitem[\protect\citeauthoryear{Gabriel et~al.,}{Gabriel
  et~al.}{2004}]{openmpi}
Gabriel E.,  et~al., 2004, in Proceedings, 11th European PVM/MPI Users' Group
  Meeting. Budapest, Hungary, pp 97--104

\bibitem[\protect\citeauthoryear{Galanti \& Roncadelli}{Galanti \&
  Roncadelli}{2018}]{galanti_behavior_2018}
Galanti G.,  Roncadelli M.,  2018, \apjm@doi [\prd]
  {10.1103/PhysRevD.98.043018}, 98, 043018

\bibitem[\protect\citeauthoryear{Giovannini, Feretti, Venturi, Kim  \&
  Kronberg}{Giovannini et~al.}{1993}]{giovannini_halo_1993}
Giovannini G.,  Feretti L.,  Venturi T.,  Kim K.~T.,   Kronberg P.~P.,  1993,
  \apjm@doi [\apj] {10.1086/172451}, 406, 399

\bibitem[\protect\citeauthoryear{Gourgouliatos, Braithwaite  \&
  Lyutikov}{Gourgouliatos et~al.}{2010}]{gourgouliatos_structure_2010}
Gourgouliatos K.~N.,  Braithwaite J.,   Lyutikov M.,  2010, \apjm@doi [Mon. Not.
  Roy. Astron. Soc.] {10.1111/j.1365-2966.2010.17410.x}, 409, 1660

\bibitem[\protect\citeauthoryear{Govoni \& Feretti}{Govoni \&
  Feretti}{2004}]{govoni_magnetic_2004}
Govoni F.,  Feretti L.,  2004, \apjm@doi [IJMPD] {10.1142/S0218271804005080}, 13, 1549

\bibitem[\protect\citeauthoryear{Govoni et~al.,}{Govoni
  et~al.}{2010}]{govoni_rotation_2010}
Govoni F.,  et~al., 2010, \apjm@doi [\aap]
  {10.1051/0004-6361/200913665}, 522, A105

\bibitem[\protect\citeauthoryear{Graham, Irastorza, Lamoreaux, Lindner  \& van
  Bibber}{Graham et~al.}{2015}]{graham_experimental_2015}
Graham P.~W.,  Irastorza I.~G.,  Lamoreaux S.~K.,  Lindner A.,   van Bibber
  K.~A.,  2015, \apjm@doi [ARNPS]
  {10.1146/annurev-nucl-102014-022120}, 65, 485

\bibitem[\protect\citeauthoryear{Guidetti, Murgia, Govoni, Parma, Gregorini, de
  Ruiter, Cameron  \& Fanti}{Guidetti
  et~al.}{2008}]{guidetti_intracluster_2008}
Guidetti D.,  Murgia M.,  Govoni F.,  Parma P.,  Gregorini L.,  de Ruiter
  H.~R.,  Cameron R.~A.,   Fanti R.,  2008, \apjm@doi [\aap] {10.1051/0004-6361:20078576}, 483, 699

\bibitem[\protect\citeauthoryear{Harari, Mollerach, Roulet  \& Sánchez}{Harari
  et~al.}{2002}]{harari_lensing_2002}
Harari D.,  Mollerach S.,  Roulet E.,   Sánchez F.,  2002, \apjm@doi [JHEP] {10.1088/1126-6708/2002/03/045}, 03, 045

\bibitem[\protect\citeauthoryear{Hardcastle}{Hardcastle}{2013}]{hardcastle_synchrotron_2013}
Hardcastle M.~J.,  2013, \apjm@doi [\mnras] {10.1093/mnras/stt1024}, 433, 3364

\bibitem[\protect\citeauthoryear{{Hitomi Collaboration} et~al.,}{{Hitomi
  Collaboration} et~al.}{2018}]{hitomi_collaboration_hitomi_2018}
{Hitomi Collaboration} et~al., 2018, \apjm@doi [\pasj] {10.1093/pasj/psx147}, 70, 13

\bibitem[\protect\citeauthoryear{Hunter}{Hunter}{2007}]{matplotlib}
Hunter J.~D.,  2007, \apjm@doi [CSE]
  {10.1109/MCSE.2007.55}, 9, 90

\bibitem[\protect\citeauthoryear{Irastorza \& Redondo}{Irastorza \&
  Redondo}{2018}]{irastorza_new_2018}
Irastorza I.~G.,  Redondo J.,  2018, \apjm@doi [PrPNP] {10.1016/j.ppnp.2018.05.003}, 102, 89

\bibitem[\protect\citeauthoryear{Irastorza et~al.,}{Irastorza
  et~al.}{2011}]{irastorza_towards_2011}
Irastorza I.~G.,  et~al., 2011, \apjm@doi [\jcap] {10.1088/1475-7516/2011/06/013}, 2011, 013

\bibitem[\protect\citeauthoryear{Kachelriess \& Tjemsland}{Kachelriess \&
  Tjemsland}{2021}]{kachelriess_origin_2021}
Kachelriess M.,  Tjemsland J.,  2021, arXiv:2111.08303 [astro-ph,
  physics:hep-ph]

\bibitem[\protect\citeauthoryear{Kalberla, Burton, Hartmann, Arnal, Bajaja,
  Morras  \& Pöppel}{Kalberla
  et~al.}{2005}]{kalberla_leidenargentinebonn_2005}
Kalberla P. M.~W.,  Burton W.~B.,  Hartmann D.,  Arnal E.~M.,  Bajaja E.,
  Morras R.,   Pöppel W. G.~L.,  2005, \apjm@doi [\aap]
  {10.1051/0004-6361:20041864}, 440, 775

\bibitem[\protect\citeauthoryear{Kale \& Parekh}{Kale \&
  Parekh}{2016}]{kale_how_2016}
Kale R.,  Parekh V.,  2016, \apjm@doi [\mnras] {10.1093/mnras/stw796}, 459, 2940

\bibitem[\protect\citeauthoryear{Kim \& Carosi}{Kim \&
  Carosi}{2010}]{kim_axions_2010}
Kim J.~E.,  Carosi G.,  2010, \apjm@doi [RvMP]
  {10.1103/RevModPhys.82.557}, 82, 557

\bibitem[\protect\citeauthoryear{Kuchar \& Enßlin}{Kuchar \&
  Enßlin}{2011}]{kuchar_magnetic_2011}
Kuchar P.,  Enßlin T.~A.,  2011, \apjm@doi [\aap]
  {10.1051/0004-6361/200913918}, 529, A13

\bibitem[\protect\citeauthoryear{Lam, Pitrou  \& Seibert}{Lam
  et~al.}{2015}]{numba}
Lam S.~K.,  Pitrou A.,   Seibert S.,  2015, in Proceedings of the Second
  Workshop on the LLVM Compiler Infrastructure in HPC. LLVM '15.
Association for Computing Machinery, New York, NY, USA,
  \apjm@doi{10.1145/2833157.2833162}

\bibitem[\protect\citeauthoryear{Libanov \& Troitsky}{Libanov \&
  Troitsky}{2020}]{libanov_impact_2020}
Libanov M.,  Troitsky S.,  2020, \apjm@doi [PhLB]
  {10.1016/j.physletb.2020.135252}, 802, 135252

\bibitem[\protect\citeauthoryear{Marsh}{Marsh}{2016}]{marsh_axion_2016}
Marsh D. J.~E.,  2016, \apjm@doi [\physrep]
  {10.1016/j.physrep.2016.06.005}, 643, 1

\bibitem[\protect\citeauthoryear{Marsh, Russell, Fabian, McNamara, Nulsen  \&
  Reynolds}{Marsh et~al.}{2017}]{marsh_new_2017}
Marsh M.~D.,  Russell H.~R.,  Fabian A.~C.,  McNamara B.~P.,  Nulsen P.,
  Reynolds C.~S.,  2017, \apjm@doi [\jcap] {10.1088/1475-7516/2017/12/036}, 12,
  036

\bibitem[\protect\citeauthoryear{Marsh et al.}{2022}]{marsh_fourier_2022} Marsh M.~C.~D., Matthews J.~H., Reynolds C., Carenza P., 2022, \apjm@doi[\prd] {10.1103/PhysRevD.105.016013}, 105, 016013. 

\bibitem[\protect\citeauthoryear{Matthews}{Matthews}{2021}]{alpro}
Matthews, J. H, 2022, alpro: Axion-Like PROpagation, v1.0, Zenodo, doi:10.5281/zenodo.6079445

\bibitem[\protect\citeauthoryear{Mckinney}{Mckinney}{2010}]{mckinney-proc-scipy-2010}
Mckinney W.,  2010, in {S}t\'efan van~der {W}alt {J}arrod {M}illman eds,
  {P}roceedings of the 9th {P}ython in {S}cience {C}onference. pp 56 -- 61,
  \apjm@doi{10.25080/Majora-92bf1922-00a}

\bibitem[\protect\citeauthoryear{McNamara, Kuncic, \& Wu}{McNamara et~al.}{2009}]{mcnamara2009} McNamara A.~L., Kuncic Z., Wu K., 2009, \apjm@doi [\mnras] {10.1111/j.1365-2966.2009.14608.x}, 395, 1507.

\bibitem[\protect\citeauthoryear{Meyer, Montanino  \& Conrad}{Meyer
  et~al.}{2014}]{meyer_detecting_2014}
Meyer M.,  Montanino D.,   Conrad J.,  2014, \apjm@doi [\jcap] {10.1088/1475-7516/2014/09/003}, 2014, 003

\bibitem[\protect\citeauthoryear{Meyer, Davies  \& Kuhlmann}{Meyer
  et~al.}{2021}]{meyer_gammaalps_2021}
Meyer M.,  Davies J.,   Kuhlmann J.,  2021, ASCL,
  p. ascl:2109.001

\bibitem[\protect\citeauthoryear{Mirizzi, Redondo  \& Sigl}{Mirizzi
  et~al.}{2009}]{mirizzi_constraining_2009}
Mirizzi A.,  Redondo J.,   Sigl G.,  2009, \apjm@doi [\jcap]
  {10.1088/1475-7516/2009/08/001}, 08, 001

\bibitem[\protect\citeauthoryear{Murgia, Govoni, Feretti, Giovannini,
  Dallacasa, Fanti, Taylor  \& Dolag}{Murgia
  et~al.}{2004}]{murgia_magnetic_2004}
Murgia M.,  Govoni F.,  Feretti L.,  Giovannini G.,  Dallacasa D.,  Fanti R.,
  Taylor G.~B.,   Dolag K.,  2004, \apjm@doi [\aap]
  {10.1051/0004-6361:20040191}, 424, 429

\bibitem[\protect\citeauthoryear{Nagai et~al.,}{Nagai
  et~al.}{2019}]{nagai_alma_2019}
Nagai H.,  et~al., 2019, \apjm@doi [\apj]
  {10.3847/1538-4357/ab3e6e}, 883, 193

\bibitem[\protect\citeauthoryear{Nandra et~al.,}{Nandra
  et~al.}{2013}]{nandra_hot_2013}
Nandra K.,  et~al., 2013, arXiv e-prints, 1306, arXiv:1306.2307

\bibitem[\protect\citeauthoryear{Payez, Evoli, Fischer, Giannotti, Mirizzi  \&
  Ringwald}{Payez et~al.}{2015}]{payez_revisiting_2015}
Payez A.,  Evoli C.,  Fischer T.,  Giannotti M.,  Mirizzi A.,   Ringwald A.,
  2015, \apjm@doi [\jcap]
  {10.1088/1475-7516/2015/02/006}, 2015, 006

\bibitem[\protect\citeauthoryear{Peccei \& Quinn}{Peccei \&
  Quinn}{1977}]{peccei_cp_1977}
Peccei R.~D.,  Quinn H.~R.,  1977, \apjm@doi [\prl]
  {10.1103/PhysRevLett.38.1440}, 38, 1440

\bibitem[\protect\citeauthoryear{Perrone \& Latter}{2021}]{perrone_2021} Perrone L.~M., Latter H., 2021, arXiv:2110.14696

\bibitem[\protect\citeauthoryear{Pfrommer \& Enßlin}{Pfrommer \&
  Enßlin}{2004}]{pfrommer_estimating_2004}
Pfrommer C.,  Enßlin T.~A.,  2004, \apjm@doi [\mnras] {10.1111/j.1365-2966.2004.07900.x}, 352, 76

\bibitem[\protect\citeauthoryear{Preskill, Wise  \& Wilczek}{Preskill
  et~al.}{1983}]{preskill_cosmology_1983}
Preskill J.,  Wise M.~B.,   Wilczek F.,  1983, \apjm@doi [PhLB]
  {10.1016/0370-2693(83)90637-8}, 120, 127

\bibitem[\protect\citeauthoryear{Raffelt}{Raffelt}{1996}]{raffelt_stars_1996}
Raffelt G.~G.,  1996, Stars as laboratories for fundamental physics : the
  astrophysics of neutrinos, axions, and other weakly interacting particles.
\url {https://ui.adsabs.harvard.edu/abs/1996slfp.book.....R}

\bibitem[\protect\citeauthoryear{Raffelt}{Raffelt}{2008}]{raffelt_astrophysical_2008}
Raffelt G.~G.,  2008, in , Vol.~741, Axions.
p.~51, \url {https://ui.adsabs.harvard.edu/abs/2008LNP...741...51R}

\bibitem[\protect\citeauthoryear{Raffelt \& Stodolsky}{Raffelt \&
  Stodolsky}{1988}]{raffelt_mixing_1988}
Raffelt G.,  Stodolsky L.,  1988, \apjm@doi [\prd]
  {10.1103/PhysRevD.37.1237}, 37, 1237

\bibitem[\protect\citeauthoryear{Reynolds, Marsh, Russell, Fabian, Smith,
  Tombesi  \& Veilleux}{Reynolds et~al.}{2020}]{reynolds_astrophysical_2020}
Reynolds C.~S.,  Marsh M. C.~D.,  Russell H.~R.,  Fabian A.~C.,  Smith R.,
  Tombesi F.,   Veilleux S.,  2020, \apjm@doi [\apj]
  {10.3847/1538-4357/ab6a0c}, 890, 59

\bibitem[\protect\citeauthoryear{Reynolds et~al.,}{Reynolds
  et~al.}{2021}]{reynolds_probing_2021}
Reynolds C.~S.,  et~al., 2021, arXiv:2108.04276 [astro-ph]

\bibitem[\protect\citeauthoryear{Ringwald}{Ringwald}{2012}]{ringwald_exploring_2012}
Ringwald A.,  2012, \apjm@doi [Physics of the Dark Universe]
  {10.1016/j.dark.2012.10.008}, 1, 116

\bibitem[\protect\citeauthoryear{Russell, Sanders  \& Fabian}{Russell
  et~al.}{2008}]{russell_direct_2008}
Russell H.~R.,  Sanders J.~S.,   Fabian A.~C.,  2008, \apjm@doi [\mnras] {10.1111/j.1365-2966.2008.13823.x}, 390,
  1207

\bibitem[\protect\citeauthoryear{Sanders \& Fabian}{Sanders \&
  Fabian}{2007}]{sanders_deeper_2007}
Sanders J.~S.,  Fabian A.~C.,  2007, \apjm@doi [\mnras]
  {10.1111/j.1365-2966.2007.12347.x}, 381, 1381

\bibitem[\protect\citeauthoryear{Sanders, Fabian  \& Dunn}{Sanders
  et~al.}{2005}]{sanders_non-thermal_2005}
Sanders J.~S.,  Fabian A.~C.,   Dunn R. J.~H.,  2005, \apjm@doi [\mnras] {10.1111/j.1365-2966.2005.09016.x}, 360,
  133

\bibitem[\protect\citeauthoryear{Schallmoser, Krippendorf, Chadha-Day  \&
  Weller}{Schallmoser et~al.}{2021}]{schallmoser_updated_2021}
Schallmoser S.,  Krippendorf S.,  Chadha-Day F.,   Weller J.,  2021,
  arXiv:2108.04827 [astro-ph, physics:hep-ph]

\bibitem[\protect\citeauthoryear{Schekochihin \& Cowley}{Schekochihin \&
  Cowley}{2006}]{schekochihin_turbulence_2006}
Schekochihin A.~A.,  Cowley S.~C.,  2006, \apjm@doi [PhPl]
  {10.1063/1.2179053}, 13, 056501

\bibitem[\protect\citeauthoryear{Schekochihin, Cowley, Kulsrud, Hammett  \&
  Sharma}{Schekochihin et~al.}{2005}]{schekochihin_plasma_2005}
Schekochihin A.~A.,  Cowley S.~C.,  Kulsrud R.~M.,  Hammett G.~W.,   Sharma P.,
   2005, \apjm@doi [\apj] {10.1086/431202}, 629, 139

\bibitem[\protect\citeauthoryear{Schekochihin et al.}{2004}]{schekochihin2004} Schekochihin A.~A., Cowley S.~C., Taylor S.~F., Maron J.~L., McWilliams J.~C., 2004, \apjm@doi [\apj] {10.1086/422547}, 612, 276

\bibitem[\protect\citeauthoryear{Schnittman \& Krolik}{2010}]{schnittman2010} Schnittman J.~D., Krolik J.~H., 2010, \apjm@doi [\apj] {doi:10.1088/0004-637X/712/2/908}, 712, 908. 

\bibitem[\protect\citeauthoryear{Sisk~Reynés, Matthews, Reynolds, Russell,
  Smith  \& Marsh}{Sisk~Reynés et~al.}{2021}]{sisk_reynes_new_2021}
Sisk~Reynés J.,  Matthews J.~H.,  Reynolds C.~S.,  Russell H.~R.,  Smith
  R.~N.,   Marsh M. C.~D.,  2022, \apjm@doi [\mnras] {10.1093/mnras/stab3464}, 510, 1264.  

\bibitem[\protect\citeauthoryear{Svrcek \& Witten}{Svrcek \&
  Witten}{2006}]{svrcek_axions_2006}
Svrcek P.,  Witten E.,  2006, \apjm@doi [JHEP]
  {10.1088/1126-6708/2006/06/051}, 2006, 051

\bibitem[\protect\citeauthoryear{Taylor, Gugliucci, Fabian, Sanders, Gentile
  \& Allen}{Taylor et~al.}{2006}]{taylor_magnetic_2006}
Taylor G.~B.,  Gugliucci N.~E.,  Fabian A.~C.,  Sanders J.~S.,  Gentile G.,
  Allen S.~W.,  2006, \apjm@doi [\mnras]
  {10.1111/j.1365-2966.2006.10244.x}, 368, 1500

\bibitem[\protect\citeauthoryear{Pandas Development Team}{Pandas Development Team}{2020}]{reback2020pandas}
Pandas Development Team,  2020, pandas-dev/pandas: Pandas,
  \apjm@doi{10.5281/zenodo.3509134}, \url
  {https://doi.org/10.5281/zenodo.3509134}

\bibitem[\protect\citeauthoryear{Tribble}{Tribble}{1991}]{tribble_radio_1991}
Tribble P.~C.,  1991, \apjm@doi [\mnras] {10.1093/mnras/253.1.147}, 253, 147

\bibitem[\protect\citeauthoryear{Ursini et al.}{2022}]{ursini2022} Ursini F., Matt G., Bianchi S., Marinucci A., Dov{\v{c}}iak M., Zhang W., 2022, \apjm@doi [\mnras] {10.1093/mnras/stab3745}, 510, 3674. 

\bibitem[\protect\citeauthoryear{Vazza, Brüggen, Gheller  \& Wang}{Vazza
  et~al.}{2014}]{vazza_amplification_2014}
Vazza F.,  Brüggen M.,  Gheller C.,   Wang P.,  2014, \apjm@doi [\mnras] {10.1093/mnras/stu1896}, 445, 3706

\bibitem[\protect\citeauthoryear{Vazza, Brunetti, Brüggen  \& Bonafede}{Vazza
  et~al.}{2018}]{vazza_resolved_2018}
Vazza F.,  Brunetti G.,  Brüggen M.,   Bonafede A.,  2018, \apjm@doi [\mnras] {10.1093/mnras/stx2830}, 474, 1672

\bibitem[\protect\citeauthoryear{Virtanen et~al.,}{Virtanen
  et~al.}{2020}]{2020SciPy-NMeth}
Virtanen P.,  et~al., 2020, \apjm@doi [Nature Methods]
  {10.1038/s41592-019-0686-2}, \href {https://rdcu.be/b08Wh} {17, 261}

\bibitem[\protect\citeauthoryear{Vogt \& Enßlin}{Vogt \&
  Enßlin}{2005}]{vogt_bayesian_2005}
Vogt C.,  Enßlin T.~A.,  2005, \apjm@doi [\aap]
  {10.1051/0004-6361:20041839}, 434, 67

\bibitem[\protect\citeauthoryear{Walker, ZuHone, Fabian  \& Sanders}{Walker
  et~al.}{2018}]{walker_split_2018}
Walker S.~A.,  ZuHone J.,  Fabian A.,   Sanders J.,  2018, \apjm@doi [Nature Astronomy] {10.1038/s41550-018-0401-8}, 2, 292

\bibitem[\protect\citeauthoryear{Weinberg}{Weinberg}{1978}]{weinberg_new_1978}
Weinberg S.,  1978, \apjm@doi [\prl]
  {10.1103/PhysRevLett.40.223}, 40, 223

\bibitem[\protect\citeauthoryear{Wilczek}{Wilczek}{1978}]{wilczek_problem_1978}
Wilczek F.,  1978, \apjm@doi [\prl]
  {10.1103/PhysRevLett.40.279}, 40, 279

\bibitem[\protect\citeauthoryear{Wouters \& Brun}{Wouters \&
  Brun}{2012}]{wouters_axion-like_2012}
Wouters D.,  Brun P.,  2012, in Boissier S.,  de Laverny P.,  Nardetto N.,
  Samadi R.,  Valls-Gabaud D.,   Wozniak H.,  eds, {SF2A}-2012: {Proceedings}
  of the {Annual} meeting of the {French} {Society} of {Astronomy} and
  {Astrophysics}. pp 637--640

\bibitem[\protect\citeauthoryear{Wouters \& Brun}{Wouters \&
  Brun}{2013}]{wouters_constraints_2013}
Wouters D.,  Brun P.,  2013, \apjm@doi [\apj]
  {10.1088/0004-637X/772/1/44}, 772, 44

\bibitem[\protect\citeauthoryear{Xu, Li, Collins, Li  \& Norman}{Xu
  et~al.}{2009}]{xu_turbulence_2009}
Xu H.,  Li H.,  Collins D.~C.,  Li S.,   Norman M.~L.,  2009, \apjm@doi [\apj] {10.1088/0004-637X/698/1/L14}, 698, L14

\bibitem[\protect\citeauthoryear{Zhuravleva et~al.,}{Zhuravleva
  et~al.}{2014}]{zhuravleva_turbulent_2014}
Zhuravleva I.,  et~al., 2014, \apjm@doi [\nat]
  {10.1038/nature13830}, 515, 85

\bibitem[\protect\citeauthoryear{de Angelis, Galanti  \& Roncadelli}{de~Angelis
  et~al.}{2011}]{de_angelis_relevance_2011}
de Angelis A.,  Galanti G.,   Roncadelli M.,  2011, \apjm@doi [\prd]
  {10.1103/PhysRevD.84.105030}, 84, 105030

\makeatother
\end{thebibliography}
\end{document}